**Shifting Beliefs, Changing Behavior: Experimental Evidence on Sanitation from India**

*Authors:*

(a) Sania Ashraf

Center for Behavior & the Environment, Rare, 1310 N Courthouse Rd, 110, Arlington, VA 22201, United States of America

ORCID ID- 0000-0001-7305-8922
(Email: asania@upenn.edu)

(ii) Cristina Bicchieri

Center for Social Norms and Behavioral Dynamics, University of Pennsylvania, Philadelphia, PA, United States

ORCID ID- 0000-0003-1648-5068
(Email: cb36@sas.upenn.edu)

(iii) Upasak Das (corresponding author)

Global Development Institute, University of Manchester, United Kingdom

ORCID ID- 0000-0002-4371-0139
(Email: upasak.das@manchester.ac.uk)

(iv) Alex Shpenev

Center for Social Norms and Behavioral Dynamics, University of Pennsylvania, Philadelphia, PA, United States

ORCID ID- 0000-0002-3739-1637
(Email: shpenev@sas.upenn.edu)

**Abstract**

*Open defecation, which is linked to poor health outcomes and lower cognitive ability has been widespread in India. Improved sanitation practice generates local health externalities, which implies that the returns to private toilet usage depend on community-wide compliance. Therefore, beliefs about others' toilet usage behavior may influence both sanitation choices and the perceived value of residing in areas with lower prevalence of open defecation. This paper evaluates a randomized, intervention in peri-urban Tamil Nadu designed to shift social expectations regarding sanitation practice. The intervention combines descriptive norm messaging, community engagement, and information provision to increase perceived prevalence and approval of toilet usage. The findings indicate a significant increase in toilet access in the intervention areas by 8 percentage points and consistent toilet usage by 6–7 percentage points relative to control areas. The effects are robust to potential bias due to disruption due to local socio-political protest and the COVID-19 pandemic during the study period. We also find significant increase in empirical expectations (beliefs about the prevalence of toilet usage) and normative expectations (beliefs about its approval). Mediation analysis indicates that the improvement in toilet access and usage is substantially driven by changes in these social expectations. We further observe significant increase in the willingness to pay among respondents for residing in open defecation free surroundings, suggestive of an increased perceived valuation of cleaner sanitation environments because of the intervention. The findings underscore the need for norm-centric interventions to propel change in beliefs and achieve long-term and sustainable sanitation behavior.*

**Acknowledgement**

*The authors thank the conference participants in Indian Statistical Institute, New Delhi; Indira Gandhi Institute of Development Research, Mumbai and Jadavpur University, Kolkata. This project was funded by the Bill and Melinda Gates Foundation (Grant No. INV-009118/ OPP1157257). The funder was not involved in study design, data collection, analysis, or interpretation. Swasti Health Catalyst implemented the intervention on the ground. Kantar Republic and Academy of Management Studies managed the data collection but had no role in the study design, decision to publish, or manuscript preparation. The study protocol has been reviewed and approved by the ethics board at the University of Pennsylvania, USA, and the Catalyst Foundation, India. Trial registration: NCT04269824. We thank Kavita Chauhan, Jinyi Kuang, Erik Thulin, Peter McNally, Maryann Delea, Rachel Sanders, and Hans Peter Kohler for their suggestions and help while setting up the project along with all the respondents and the investigators of the survey work. We also thank Anubha Tyagi and Paulius Yamin for their logistical assistance.*

## 1. Introduction

Inadequate access to improved sanitation is linked to soil-transmitted helminths, enteric diseases, infant mortality, reduced physical and cognitive growth among children, and poorer educational outcomes (Saleem et al., 2019; Strunz et al., 2014; Hammer and Spears, 2016; Spears and Lamba, 2016; Geruso and Spears, 2018; Spears, 2020; Cameron et al., 2021). The Sustainable Development Goals' (SDGs) call for universal access to sanitation and ending open defecation (OD), yet India faces significant challenges in providing safe sanitation to its population. Despite government-led nationwide sanitation programs, such as the Swachh Bharat Abhiyan (SBA), there has been mixed success in promoting consistent toilet usage. While some studies have documented evidence of the impact of the program on toilet access, its use remains low in India (Barnard et al., 2013; Coffey et al., 2017). Recent data from the National Family Health Survey (2019-21) also indicates that 17 percent of households still do not have access to a toilet, with rural areas reporting up to 25 percent.[1]

One of the striking features of OD practice in India is that it is not always perceived as a health threat, and in some cases is viewed as "pleasurable, comfortable, or convenient" (Coffey et al., 2014). In fact, studies have documented considerable prevalence of OD even among those with access to toilets (Bicchieri et al., 2018). Literature also have noted significant slippage in toilet usage over time, with individuals resorting to defecating in the open while using toilets before (Coffey et al., 2017; Abebe and Tucho, 2020). These patterns raise an important question: beyond prices and access constraints, do beliefs about others' toilet usage behavior influence one's sanitation choices? This paper utilizes a randomly implemented intervention based on influencing social beliefs in peri-urban areas of Tamil Nadu, India, to assess its effect on toilet access and usage behavior. In the process, we also assess the effects on one’s own expectations around sanitation practices in the community, and valuation of areas with lower prevalence of OD.

[1] Details on the National Family Health Survey data can be obtained from https://www.nfhsiips.in/nfhsuser/index.php (accessed on March 202, 2026)

The intervention is based on the idea that people's beliefs and behaviors are often influenced by their perception of community behavior and approval. In other words, one's belief about the prevalence of certain behavior within the community and its approval among its members may motivate them to behave pro-socially (Reno et al. 1993; Frey and Meier, 2004; Bicchieri, 2006; Bicchieri 2016; Steg and Vlek, 2009, Krupka and Weber, 2009). Therefore, these behaviors are influenced through social expectations, which might take the form of Empirical Expectations (EE) (expectations regarding what others around them do) or Normative Expectations (NE) (expectations regarding what others around them think one should do) (Bicchieri, 2006; Bicchieri, 2016). Because sanitation generates local health externalities (Augsburg et al. 2024), the marginal health return to private toilet usage depends on community-wide compliance. Hence, if beliefs about prevalence of toilet usage and approval in the community increase, it may lead to increase in the likelihood of improved latrine usage. When environmental contamination declines only if a sufficiently large share of households adopt improved practices, individual incentives to invest in and use toilets depend on beliefs about others’ behavior. If individuals believe that only few use toilets or approve its usage, they may perceive limited environmental cleanliness and low marginal returns to private investment, thereby reducing incentives to construct or consistently use toilets. An increase in EE raises the perceived prevalence of toilet usage, thereby increasing the expected cleanliness of the local environment and the marginal health benefit of private compliance. In addition, as compliance becomes more prevalent and most households use toilets and disapprove open defecation, deviance becomes more visible and socially costly, increasing the expected reputational penalty associated with non-compliance (Azar, 2004; Miguel & Gugerty 2005).

Early diagnostic research using non-experimental data on OD across rural and urban parts of Bihar and Tamil Nadu in India indicates that toilet usage is highly associated with EE and, to a lesser extent, with NE (Bicchieri et al., 2018). In other words, individuals are more likely to use a toilet for defecation if they believe that others in the community also use a toilet and approve of this practice. Building on this research, we developed an intervention based on social expectation that involved messages and activities aiming to increase the EE and NE of the community members surrounding toilet usage, while additionally providing them with relevant information on toilet construction and usage. Specifically, there were four components of the intervention. The first component was community mobilization and events where influential political and community leaders applaud toilet users and promote improved sanitation practices. The second component was mass media broadcasting information on the actual and/or prospective adoption of improved sanitation practices, while the third component involved peer-counselling sessions to share information on toilet construction and access to finance. The final component was household visits by outreach workers to discuss sanitation practices and then use stickers (thumbs up signs) to signal improved sanitary practices. The target is to shift social beliefs about other people's latrine usage, which can motivate people to construct and use toilets for defecation.

The intervention was randomly rolled out in 38 out of the 76 eligible wards across five town panchayats (peri-urban towns) located in two districts of Tamil Nadu. The remaining 38 wards are used control units. In this paper, we assess the impact of this intervention on two broad outcomes: toilet access in the household and exclusive toilet usage among individuals. To understand the mechanisms, we explore whether the intervention led to an increase in social expectations surrounding toilet usage behavior within their community, namely one’s EE and examine whether the effects on toilet access and usage are mediated by this increase. Further, because improved sanitation practices can generate local positive externalities, it is possible that the potential increase in perceived community toilet usage and its approval can increase the expected marginal health return to residing in areas with cleaner sanitation environments. Accordingly, we also assess the impact of the intervention in altering in valuation for residing in surroundings with low prevalence of OD.

Because of the COVID-19 pandemic, the study was not able to survey the same households in the endline as those in the baseline. Therefore, it employs a repeated cross-section design with data collected from 2571 households in the baseline and then in the endline as well. The findings indicate that the intervention led to a significant increase in the likelihood of toilet access by 8 percentage points (over control mean of 45 percent) and exclusive toilet usage by 6-7 percentage points (over control value of about 70 percent). We also observe that the intervention led to a significant increase in perception of the prevalence of toilet usage and ownership in the community (EE). Additionally, more people started approving toilet usage (NE). The results are robust to bias-adjusted treatment effects and also non-random sample selection due to socio-political events like the protests against the Citizenship Amendment Act (CAA) and National Register of Citizens (NRC). While there can be other mechanism at work, we find that the impact of the intervention on toilet access and usage is explained by the associated changes in EE and NE. Furthermore, the regression estimations also indicate about an improvement in perception or valuation of residing in open defecation free (ODF) and low OD areas elicited through the WTP.

The paper offers several contributions. First, our paper contributes to the literature on interventions that promote health behavior. Broadly, there are two types of interventions to achieve this. The first one relies on free distribution, loans or subsidies that lower the cost of healthy behavior and enhance adoption of health products (Dupas, 2014; Tarozzi et al., 2014; Gertler, 2015). Additionally, in poorer settings, subsidies might be essential to ensure widespread adoption of healthy behavior including sanitation investments (Cohen and Dupas, 2010; Guiteras et al. 2015; Gautam et al. 2025). The second type of intervention relies on behavioral change interventions, depending on norm messaging and nudging, which also can lower the cost of adhering to pro-social behavior (Ashraf et al., 2006; Thaler and Sunstein, 2009; Giné et al., 2010). Related to this are also interventions linked with reputational concerns and social image (Benabou and Tirole, 2003; Bénabou and Tirole, 2006) that have been found to have persistent impact (Bakhtiar et al. 2023). This paper contributes to studies that look at the second type, namely behavioral change interventions to promote healthy behavior.

Second, the paper also adds to the growing literature on norms and its application in field experiments. While existing literature in Economics has focused on the persistence of cultural traits and norms (Giuliano, 2007; Alesina et al., 2013), we study how the provision of information (norm messaging) on behavior prevalence can potentially change existing beliefs and behaviors. To study the effects, studies have often utilized lab-based experiments or randomized hypothetical vignettes (Krupka and Weber, 2013; Bicchieri et al., 2022a; Bicchieri et al., 2022b). Field-based experimental studies that revolve around norm-centric interventions are limited and this paper is a contribution to specific literature.

Third, our paper examines how behavioral change and psychological interventions influence the overall change in preferences and expectations. Prior research has demonstrated that norm messaging effectively reduces the prevalence of drug use, sexual assaults, and drinking problems (Perkins, 2003; Kramer and Levy, 2008; Hillenbrand-Gunn et al., 2010). Similarly, psychological interventions have been shown to improve preventive health investments (John and Orkin, 2021). In this paper, we evaluate the impact on toilet usage and individual valuation of open defecation free surrounding. This is in line with previous existing literature that has looked into willingness to pay for better environment quality (Yishay et al. 2017). This is particularly relevant in the context of sanitation behavior, where the way individuals value improved sanitation within their community is key to imposing sanctions and maintaining ODF surroundings that can ensure long-term sustenance.

Fourth, previous studies that examined norm-based messaging to promote latrine usage in India have predominantly been guided by approaches such as Behavior Centered Design (BCD), Community-Led Total Sanitation (CLTS), and the Risks, Attitudes, Norms, Abilities, and Self-regulation (RANAS). While these approaches recognize social norms as drivers of behavior change, they do not explicitly leverage specific social expectations that can effectively influence norms (Aunger and Curtis, 2016; Kar and Chambers, 2009; Friedrich et al., 2020). In contrast, our intervention stands out as among the first to primarily focus on altering social expectations to achieve improved sanitation practices.

In terms of policy implications, this study holds significant relevance in the context of ensuring healthy environments and better health outcomes, particularly in the post-COVID scenario. Our findings contribute to the design of effective norm-centric interventions for addressing open defecation, a persistent challenge in India and other low- and middle-income countries. By offering insights into how norm-based interventions can drive sustained behavioral change, our study provides valuable guidance for policymakers and practitioners working on improving sanitation and other healthy behaviors.

The paper is structured as follows. Section 2 discusses the motivation behind the intervention and its components. Section 3 explains the randomization process and data used in the analysis. Section 4 describes the empirical strategy and section 5 presents the estimations from the main regressions, robustness checks, and mediation analysis. Section 6 discusses some other further analyses and Section 7 concludes with a discussion.

**2. Intervention- motivation and design**

Our intervention revolves around the fundamental question: Does an individual's belief about others within their community influence their own beliefs and actions? Previous studies have indicated that beliefs about how individuals are perceived by others play a significant role in shaping their decisions and behavior (Della Vigna et al., 2012; Bursztyn and Jensen, 2015; Bursztyn et al., 2017; Della Vigna et al., 2017). This question becomes particularly relevant for behaviors which are deeply rooted in community life. Therefore, it is possible that interventions focusing on community behaviors and beliefs (non-factual beliefs targeting both the individual and community levels) can have the potential to motivate people to engage in prosocial behaviors.

Our intervention is grounded in the idea that behaviors are often interdependent and thus influenced by social expectations regarding the actions or the endorsements of community members (Cialdini et al., 1990; Bicchieri, 2006; Bicchieri, 2016). Social expectations can be EE reflecting an individual's beliefs about the prevalence of a behavior within the community, or NE, encompassing beliefs about the approval of that behavior within the community. Experimental and non-experimental studies conducted in Economics and other disciplines have demonstrated that interventions that provide information on the actions or beliefs of others can be used to bring in behavioural change in several domains, such as higher charitable contributions, retirement savings decisions, voting behavior, female labor participation, exclusive breastfeeding and promoting energy conservation (Gerber and Rogers, 2009; Shang and Croson, 2009; Beshears et al. 2015; Bonan et al. 2020; Bursztyn et al., 2020; Bicchieri et al., 2022a). In the context of sanitation, utilizing non-experimental data and randomly assigned experimental vignettes, it has been shown that individual perception about prevalence of toilet usage within the community is a strongly predictor of whether the individual defecates in the open or not (Bicchieri et al., 2018; Thulin et al., 2021).[2]

Utilizing the findings from these studies, we develop our norm-centric intervention, called the *Nam Nalavazhvu* (NN) for implementation in peri-urban areas of Tamil Nadu.[3] This involved a multistep and systematic process, including problem and solution tree analysis, intervention mapping, and identification of potential intervention techniques (Starr, 2017). We utilized the results to map the behavioral factors influencing OD in our target population and outlined a set of potential intervention techniques guided by Michie et al. (2011). It is important to note that our study was implemented shortly after a significant increase in toilet coverage through the SBA, and most states had been declared ODF by the government. We leveraged this context to disseminate information about improved toilet usage in the community, aiming to shift EE through descriptive norm messaging (Cialdini et al., 2006). By providing relevant information on improved sanitation practices, our intervention aimed to potentially encourage others to conform to the new norm of toilet use.

The socio-behavioral communication materials highlighted messages that were designed to generate a call to action, emphasizing improved toilet usage and signaling positive changes within the community. The intervention operated at both, ward levels as well as household levels within each intervention cluster (ward in our case). At the ward level, community mobilization and commitment events were organized through the involvement of political and community leaders, aimed at acknowledging toilet users and promoting improved sanitation practices. Audio broadcasts were conducted using automobiles, and wall paintings were utilized to promote toilet ownership and usage. Influential community members were also engaged to disseminate promotional messages related to improved sanitary practices through social media networks.

[2] Extensive formative research using longitudinal surveys, and focus group discussions across rural, peri-urban and urban slums of Bihar and Tamil Nadu have been carried out prior to the intervention to understand the social determinants of toilet usage (Bicchieri et al. 2017; Bicchieri et al. 2018; Ashraf et. al 2022). Please note that the districts where these studies have been conducted are different from the ones, which are intervened and studied in this paper.

[3] "Nam Nalavazhvu" in Tamil language means "Our Well-Being"

At the household level, visual stickers were used to indicate households that adhered to improved sanitary practices. In addition to the descriptive norm messaging, we provided household-level information on resources and various toilet designs to ensure that the created demand could be effectively channelled. Outreach workers utilized flipbooks containing information on financial resources and toilet designs, enabling interested individuals to connect with relevant resources within their community for toilet construction. Peer-counselling sessions were organized in small groups with neighbors to leverage social networks, facilitate conversations, and provide information on access to sanitation markets, technologies, and barriers.[4] Further details of the intervention components can be found in Ashraf et al. (2021).

In summary, the intervention had two broad dimensions: descriptive norm messaging and the provision of information and facilitated discussions on financing, toilet technology, and markets. The goal was to shift social expectations within the community by increasing perceived prevalence of toilet usage and indirectly its approval (EE and NE). Theoretically it can raise the expected health returns to private compliance and reduces coordination failures that may sustain low-adoption equilibria. In addition, greater visibility of adoption may induce informational learning about the benefits of sanitation and strengthen informal enforcement through social approval and disapproval. Through these mechanisms, the intervention is expected to increase both toilet access and consistent usage, and to enhance the value individuals place on residing in environments with improved sanitation. It is important to note that we conducted the trial of improved practices for three months in the same districts but in a different town panchayat distant from the study areas.[5] These trials allowed us to refine and revise the activities that were finally implemented.

## 3. Randomization process and data

[4] For detailed information on the intervention, refer to Ashraf et al. (2021). We have also laid some details in Online Appendix A1 too.

[5] The peri-urban areas within a district are divided into 3-4 Town Panchayats in Tamil Nadu. These Town Panchayats are broader administrative units in the peri-urban areas.

As discussed, the NN intervention was randomly implemented across peri-urban areas in the Pudukottai and Karur districts in the state of Tamil Nadu, India. Specifically, we selected five town panchayats in each of the two districts.[6] The locations of Tamil Nadu, along with these two districts, are shown in Figure 1. Within these town panchayats are clusters called wards, which form the unit of randomization. To identify the potential wards for our study, we sought assistance from the local executive officers through whom we obtained the ward maps of each town panchayat. Wards that were primarily commercialized with very few residential households were excluded from our study. Additionally, we excluded urbanized wards that had complete toilet access coverage according to official records.[7] We also excluded wards located on the borders of two or more adjoining wards to minimize contamination between treatment and control wards. It is important to note that we ensured a minimum distance of 1 km between each of the non-excluded wards.

Figure 1: Location of Tamil Nadu and Pudukottai and Karur districts within Tamil Nadu

[6] The three town panchayats, which were used for piloting the intervention, have been excluded from the sampling frame. Further, we had to drop another town panchayat from which we could not receive permission to study because of political concerns.

[7] These wards are typically those where government officials or school/ college teachers live. Some of these wards also have factories in which factory workers live in allocated housing complexes. Every apartment in such complexes has been provided with improved toilets and hence OD is negligible in these wards.

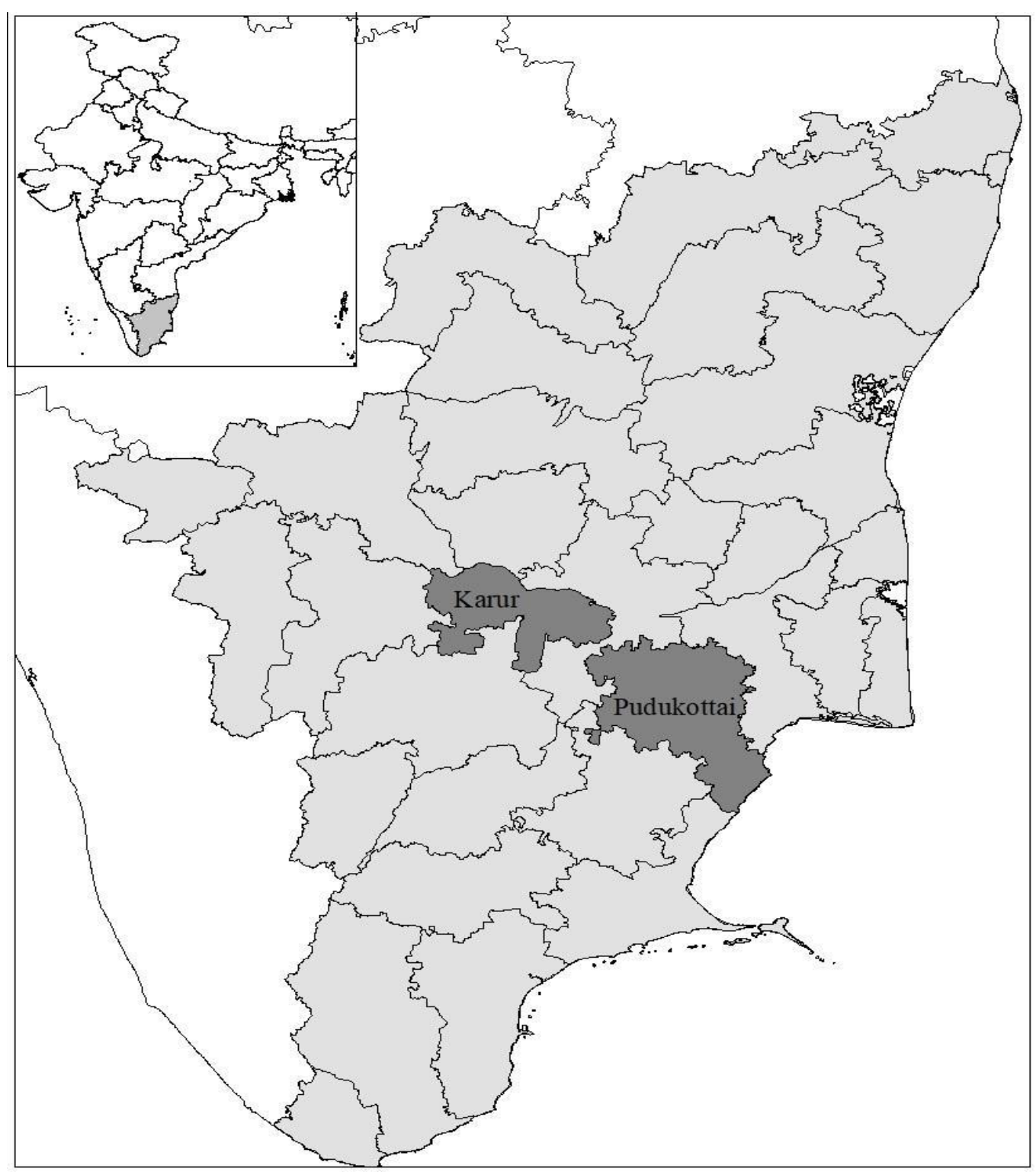

Note: This is the map of Tamil Nadu with the districts. The darker shaded areas are the districts of Karur and Pudukottai. In the inner box, the map of India is presented with the shaded area being the state of Tamil Nadu.

In the ten randomly selected town panchayats, 79 eligible wards were identified based on the aforementioned criteria. From these 79 wards, a sample of 76 wards was randomly chosen for our study, aiming to represent peri-urban wards primarily used for residential purposes. Out of these 76 wards, 38 wards were randomly assigned to receive the NN intervention, making them the treatment wards, while the remaining wards served as controls. All households residing in the treatment wards were eligible to participate in the activities mentioned above.

In the baseline survey, we gathered data from 2,571 households and our original plan was to re-survey the same households in the endline as well. However, due to various factors primarily related to the COVID-19 pandemic such as migration, illness, and the non-entry of surveyors, we tried to but could not necessarily reach the same households during endline.[8] Therefore, in our analysis, we treat the surveyed 2,571 households in the endline as an independent cross-section and use the baseline and the endline household data as repeated-cross section. Other studies conducted during the pandemic also followed a similar strategy (Chimbutane et al. 2026).

[8] We had to change the survey agency as well from baseline to the endline because of the pandemic. However, we ensured that the instructions given to the survey agencies were similar.

The survey questionnaire collected a wide range of information, including household socio-demographic and economic characteristics. We also gathered detailed data on household toilet ownership, respondents' defecation practices, and their social expectations about toilet usage. Specific questions on willingness to pay for residing in ODF or low OD areas were asked. The endline survey also collected information on the extent of exposure of the respondents to the intervention across the treatment and control wards.

**Variables**

Our main outcome variables are toilet access and usage, for which we use the following questions asked during the two rounds of survey*:*

(i) "Do members of your household have access to a private, public, or community toilet?"

(ii) "Some people defecate in the open and some people use a toilet. Where did you defecate the last time you needed to?"

(iii) "During the last two days, where was your primary place of defecation?"

(iv) "During the last two days, did you only use a toilet for defecation?"

(v) "During the last seven days, including today, did you use a toilet for defecation when you needed to defecate?".

The first question is used to study the impact on toilet access and questions (ii) to (v) are used to assess the impact of the intervention on toilet usage. Please note that the questions (iii), (iv) and (v) are used to capture whether toilets have been used exclusively for defecation purpose in the last two days or past one week.[9] All these five questions are converted to binary indicators that reflect access to private toilet and exclusive usage of toilet and open defecating in the open.

[9] If the respondent reports "everytime" to the question "During the last seven days, including today, did you use a toilet for defecation when you needed to defecate?", it can be reasonably assumed that toilet is used exclusively for defecation purpose.

To explore the mechanisms, we mainly consider two aspects: EE and NE. To measure EE, we use the following four variables:

(i) "Out of ten members in your ward, how many do you think use a toilet every time they need to defecate?";

(ii) "Out of ten households in your ward, how many do you think own a toilet at home?";

(iii) "Many people in my ward built new toilets in the past six months"; and

(iv) "I think more people expect others to use toilets compared to six months ago".

These questions elicit what individuals believe about sanitation practices- belief about the prevalence (Bicchieri, 2022) and have been used in other contexts (Bicchieri et al. 2022a; Babbar et al. 2023). Important here is to note that before asking the questions on defecation practice, we precede by a "statement of normalization" which is as follows: "Some people defecate in the open and some people use a toilet." This statement intends to reduce the likelihood of respondents to feel obligatory pressures in responding in a certain way thereby potentially reducing the social desirability bias (Bicchieri et al. 2022).

Response to question (i) and (ii) varies from 0 to 10 and the response value is used directly in the regression to gauge the treatment effects on EE. For question (iii) and (iv), we gave five options: strongly disagree; disagree; neither; agree; strongly agree. We convert them in binary form wherein we assign a value of 1 if the respondents report agree/ strongly agree and 0 otherwise. For NE, we use *"Out of ten members of your ward, how many do you think believe one should use a toilet to defecate?"*. As in the case for (i) and (ii), we use the response value directly in the regression to estimate the impact on NE. The definition of all the indicators is presented in Appendix Table 2.

**4. Estimation Strategy**

We utilize the experimental design of our intervention to generate the unbiased causal estimates. Since we use repeated cross-section data, we do not have baseline information for households surveyed in the endline. Therefore, we use the Analysis of Covariance (ANCOVA) framework to control for the mean baseline values at the ward level for estimating the treatment effects (McKenzie, 2012; Hidrobo et al.

2016; Haushofer et al. 2020, Das et al. 2023b) similar to what has employed by Chimbutane et al. (2026). To assess the impact on toilet access and usage, we estimate the following regression equation:

$$Y_{iwpt=1} = \alpha + \beta.Treatment_{wp} + \gamma Mean(Y_{iwpt=0}) + \delta X_{iwpt=1} + \pi_p + \varepsilon_{iwdt=1} \quad (1)$$

Here, $t = 1$ indicates values from the endline wave and $t = 0$ is the baseline. $Y_{iwpt=1}$ is the indicator of toilet access and usage as reported by respondent, $i$ residing in ward, $w$ located in town panchayat, $p$ at endline. The treatment status of ward, $w$, is indicated by $Treatment_{wp}$, wherein the variable takes the value of 1 if it has been intervened and 0 otherwise. $X_{iwpt=1}$ is the vector of individual and household level characteristics for those surveyed in the endline and the town panchayat fixed effects is indicated by $\pi_p$. The error term is given by $\varepsilon_{iwd}$. The standard errors are clustered at the ward level. To assess the effects on EE and NE, we use the same estimation strategy.

To ensure the success of randomization and hence the causal estimates are unbiased, we consider several household and individual-level characteristics and compared the treatment and control groups based on their baseline values. All the indicators show similar mean values across the two groups, as presented in Table 1 and none was 5 percent level of significance.[10] As stated in the pre-analysis plan, we construct a household asset index using principal component analysis (PCA) by considering possession of assets such as color TV, mobile phone, motorcycle/scooter, fridge, internet, computer, AC/cooler, and washing machine.[11] We compare the asset index values between households in the treated and control wards. Additionally, we compare toilet ownership and toilet usage for defecation in the last time, three days, and seven days prior to the survey. We also assess the differences in EE and NE, along with other measures of social expectations. For the all the variables, we do not find any significant differences between the treatment and control groups of wards at 5 percent level of significance. This provides evidence in support of the randomization process we implemented and ensures that the estimates obtained from equation 1 should be unbiased.

[10] The definition of the variables has been outlined in Online Appendix A2.

[11] The pre-analysis plan has been documented here: https://cdn.clinicaltrials.gov/large-docs/24/NCT04269824/SAP_000.pdf (accessed on March 202, 2026)

Table 1: Balance between treatment and control arms (baseline values)

| | Control | | Treatment | | |
|---|---|---|---|---|---|
| | Observations | Mean/ Proportion | Observations | Mean/ Proportion | Difference |
| | | *Covariates* | | | |
| Age | 1280 | 44.894 | 1291 | 44.967 | -0.073 |
| Years of education | 1280 | 7.908 | 1291 | 7.954 | -0.046 |
| Gender: female | 1280 | 0.536 | 1291 | 0.508 | 0.028 |
| Currently married | 1280 | 0.709 | 1291 | 0.726 | -0.016 |
| Upper caste | 1278 | 0.182 | 1290 | 0.198 | -0.016 |
| Household size | 1280 | 3.623 | 1291 | 3.577 | 0.046 |
| Owns color TV | 1280 | 0.928 | 1291 | 0.920 | 0.008 |
| Has internet connection | 1279 | 0.460 | 1289 | 0.476 | -0.017 |
| Owns computer | 1280 | 0.055 | 1290 | 0.068 | -0.013 |
| Owns air-conditioner | 1279 | 0.029 | 1287 | 0.054 | -0.025* |
| Owns washing machine | 1279 | 0.120 | 1290 | 0.152 | -0.032 |
| Owns mobile phone | 1279 | 0.934 | 1291 | 0.931 | 0.003 |
| Owns motorcycle | 1279 | 0.690 | 1291 | 0.710 | -0.019 |
| Owns refrigerator | 1279 | 0.470 | 1291 | 0.497 | -0.027 |
| Owns car | 1278 | 1.967 | 1287 | 1.943 | 0.025* |
| Owns dish antenna | 1280 | 1.788 | 1288 | 1.792 | -0.004 |
| Asset index (pre-analysis plan) | 1274 | -0.115 | 1283 | 0.012 | -0.127 |
| Gas as cooking fuel | 1280 | 0.760 | 1291 | 0.813 | -0.053 |
| Has separate kitchen | 1280 | 0.738 | 1291 | 0.761 | -0.022 |
| Occupation: laborer | 1280 | 0.200 | 1291 | 0.209 | -0.009 |
| | | *Social Expectations* | | | |
| EE: How many out of 10 people use a toilet every time to defecate? | 1273 | 7.256 | 1276 | 7.611 | -0.354 |
| EE: Many people in my area are constructing a toilet | 1269 | 0.863 | 1283 | 0.832 | 0.031 |
| NE: How many out of 10 people think one should use a toilet to defecate? | 1276 | 8.738 | 1280 | 8.909 | -0.170 |
| *Toilet ownership and usage* | | | | | |
| All members have access to toilet | 1280 | 0.777 | 1290 | 0.821 | -0.044 |
| Last usage: toilet | 1280 | 0.741 | 1291 | 0.779 | -0.039 |
| Last two days primary defecation place: toilet | 1280 | 0.737 | 1291 | 0.772 | -0.036 |
| Exclusive toilets usage in last two days | 1280 | 0.665 | 1291 | 0.706 | -0.042 |
| Exclusive toilets usage in last seven days | 1280 | 0.657 | 1291 | 0.702 | -0.045 |

Mean/ proportion are reported. The difference in mean levels/ proportion is tested through the t-test. Full sample includes all households surveyed in the baseline. Full definition of these variables is given in online Appendix table A2. * denotes significance at the 10%, ** at the 5%, and *** at the 1% levels.

## Results

*Main outcome indicators: effect on toilet ownership and usage*

Table 2 presents the estimates of the intervention effect on the outcome variables discussed in the Section 3, as described in Equation 1. We run the regressions using Linear Probability Model (LPM) and present the Intent to Treat (ITT) estimate of being in the treated ward. We estimate two specifications: the first one does not control for any covariate, and the second one includes individual and household characteristics (as listed in Online Appendix A2), along with the town panchayat fixed effects. The results from the first specification are presented in columns (1) and (5) and those from the second one, which is also our preferred specification, are presented in columns (6) and (10) of Table 2.

The findings indicate that the intervention had a statistically significant 6-7 percentage points increase in the likelihood of using toilet for defecation purpose. This is over the mean level of 69-70 percent using the toilet that we find in the control wards. Importantly, exclusive toilet usage for defecation purposes is found to be higher when we use two-day and particularly the seven-day period. In control wards, we find 64 percent to use toilet exclusively in the past seven days. Over this, the intervention led to an improvement by 7 percentage points. Further, we also find that the likelihood of all household members having access to latrines increases by 8 percentage points in the treated wards relative to the control wards, where the share of households with toilet access is about 45 percent. We run a number of robustness checks that include results using the controls outlined in the pre-analysis plan, bias adjusted treatment effects, inverse probability weighing and excluding wards affected by the CAA-NRC protest.[12] Overall, our results allow us to confirm about significant treatment gains in toilet access and usage as well (Appendix R).

[12] The CAA 2019 was a law passed by the Indian government to provide a faster pathway to Indian citizenship for certain undocumented migrants specifically Hindus, Sikhs, Buddhists, Jains, Parsis, and Christians from Pakistan, Bangladesh, and Afghanistan who had entered India before 2015. The controversy arose because the law explicitly excluded Muslims. The concerns intensified when discussed alongside the proposed National Register of Citizens (NRC), a process intended to create an official list of Indian citizens. Many of the areas experienced heightened political tensions, restrictions on movement, and intermittent disruptions to public life. In light of these conditions, we were unable to safely and reliably conduct surveys in such locations

Table 2: Effect on toilet access and usage

| | Last usage: toilet | Last two days primary defecation place: toilet | Exclusive toilets usage in last two days | Exclusive toilets usage in last seven days | All members have access to toilet | Last usage: toilet | Last two days primary defecation place: toilet | Exclusive toilets usage in last two days | Exclusive toilets usage in last seven days | All members have access to toilet |
|---|---|---|---|---|---|---|---|---|---|---|
| | (1) | (2) | (3) | (4) | (5) | (6) | (7) | (8) | (9) | (10) |
| Treatment | 0.104** | 0.105** | 0.104** | 0.113** | 0.124** | 0.064** | 0.067** | 0.064** | 0.067** | 0.081* |
| | (0.048) | (0.048) | (0.048) | (0.049) | (0.054) | (0.029) | (0.029) | (0.029) | (0.028) | (0.042) |
| All controls | No | No | No | No | No | Yes | Yes | Yes | Yes | Yes |
| Baseline mean at the ward-level | No | No | No | No | No | Yes | Yes | Yes | Yes | Yes |
| Town Panchayat FE | No | No | No | No | No | Yes | Yes | Yes | Yes | Yes |
| Mean/ proportion in control wards (endline) | 0.697 | 0.694 | 0.688 | 0.641 | 0.453 | 0.697 | 0.694 | 0.688 | 0.641 | 0.453 |
| Observations | 2,570 | 2,571 | 2,571 | 2,571 | 2,571 | 2,561 | 2,562 | 2,562 | 2,562 | 2,562 |

Note: The marginal effects from the LPM model are presented. The standard errors clustered at the ward level are presented in the parenthesis. The controls include age of the respondent, years of education, gender, marital status, caste affiliation, household size and occupation along with possession of assets listed in Table 1, usage of LPG as a cooking fuel and using a separate room as kitchen. Town panchayat Fixed Effects (FE) are also included in the model. * denotes significance at the 10%, ** at the 5%, and *** at the 1% levels. The regression table with all the covariates is given in Online Appendix table A3.

*Mechanisms: The role of EE and NE*

In this section, we explore the role of empirical expectations surrounding improved sanitation practices, which forms the crux of the intervention. If the intervention is able to increase EE and individuals perceive that many within the community are adopting toilets and using them, they revise the expected cleanliness of the local environment upwards, and the marginal returns to toilet construction and consistent usage increase. Further when toilet usage becomes widespread, deviation can become more visible and potentially more socially costly. This may increase the incentives for individuals to use the latrines for defecation purpose.

To study this, we first examine if the intervention leads to a significant increase in EE. Because it is possible the perceived approval of toilet usage also can increase because of the associated increase in perceived usage, we also look at whether the intervention is successful in altering the NE as well. The variables we use to capture NE and EE are explained in Section 3 and outlined in Appendix Table A2. The estimates from the regression as specified in equation 1 using these indicators of EE and NE as outcome variables are presented in Table 3. Please note that for two indicators, "How many out of 10 own a toilet?" and "Many people are using a toilet", we did not collect baseline data and hence we are unable to control for the ward-level mean at the baseline. We observe a significant increase in both, EE and NE on average because of the intervention. Not only are the respondents from the intervention wards more likely to report higher toilet usage in their community but also, they think that more households around them are building toilets. In addition, we also find that they are more likely to think that their community approves of toilet usage for defecation.

Table 3: Effect on social expectations

| | Empirical expectations | | | | Normative expectations |
|---|---|---|---|---|---|
| | How many out of 10 use a toilet to defecate? | Many people are constructing a toilet | How many out of 10 own a toilet? | Many people are using a toilet | How many out of 10 think one should use a toilet to defecate? |
| Treatment | 0.054 | 0.125*** | 0.666*** | 0.125*** | 0.625*** |
| | (0.285) | (0.035) | (0.192) | (0.034) | (0.224) |
| All controls | Yes | Yes | Yes | Yes | Yes |
| Baseline mean at the ward-level | Yes | Yes | No | No | Yes |
| Mean/ proportion in control wards (endline) | 8.620 | 0.687 | 7.077 | 0.725 | 7.278 |
| Observations | 2,562 | 2,562 | 2,562 | 2,562 | 2,562 |
| R-squared | 0.068 | 0.077 | 0.353 | 0.095 | 0.214 |

Note: The marginal effects from the LPM model are presented. The standard errors clustered at the ward level are presented in the parenthesis. The controls include age of the respondent, years of education, gender, marital status, caste affiliation, household size and occupation along with possession of assets listed in Table 1, usage of LPG as a cooking fuel and using a separate room as kitchen. Town panchayat Fixed Effects (FE) are also included in the model. * denotes significance at the 10%, ** at the 5%, and *** at the 1% levels.

Next, we use standard regression analysis to explore the relevance of EE and NE in explaining the average treatment effects as channels. For this, we consider all those indicators of EE and NE as outlined in table 3 and incorporate them one by one in our main regression along with the other covariates and examine the changes in the marginal effects of the outcome variables when compared to the regression estimates without these variables. Table 4 presents the estimations for toilet access and exclusive toilets usage in the last seven days. The findings reveal a substantial drop in the marginal effects of the treatment for both these outcomes. Importantly, when all the indicators of EE are controlled together, we find that the effect of the intervention on toilet access and exclusive usage to be negligible. When the NE indicator ("How many out of 10 think one should use a toilet to defecate?") is controlled for, we observe a drop in the effect size of toilet access from 8.1 to 7.4 percentage points. For exclusive seven-day toilet usage, the drop in the treatment effect is about 2.3 percentage points (6.8 to 4.5 percentage points). We find similar results for the other indicators of toilet usage (Appendix M).

Table 4: Effects after controlling for EE and NE (Exclusive toilet usage in last seven days and access)

| | No EE or NE controls | Controlling for EE | | | | | Controlling for NE | Controlling for EE and NE |
|---|---|---|---|---|---|---|---|---|
| *Toilet access: All members have access to toilet* | | | | | | | | |
| Treatment | 0.081* | 0.082* | 0.075* | 0.076* | 0.070* | 0.063 | 0.074* | 0.061 |
| | (0.042) | (0.041) | (0.042) | (0.042) | (0.041) | (0.041) | (0.043) | (0.042) |
| Controls additionally added in the regression model | | | | | | | | |
| How many out of 10 use a toilet? | No | Yes | No | No | No | Yes | No | Yes |
| How many out of 10 own a toilet? | No | No | Yes | No | No | Yes | No | Yes |
| Many people are constructing a toilet (agree) | No | No | No | Yes | No | Yes | No | Yes |
| Many people are using a toilet (agree) | No | No | No | No | Yes | Yes | No | Yes |
| How many out of 10 think one should use a toilet to defecate? | No | No | No | No | No | No | Yes | Yes |
| *Exclusive toilets usage in last seven days* | | | | | | | | |
| Treatment | 0.067** | 0.066** | 0.044 | 0.049* | 0.042 | 0.031 | 0.044 | 0.029 |
| | (0.028) | (0.027) | (0.028) | (0.027) | (0.025) | (0.026) | (0.027) | (0.026) |
| Controls additionally added in the regression model | | | | | | | | |
| How many out of 10 use a toilet? | No | Yes | No | No | No | Yes | No | Yes |
| How many out of 10 own a toilet? | No | No | Yes | No | No | Yes | No | Yes |
| Many people are constructing a toilet (agree) | No | No | No | Yes | No | Yes | No | Yes |
| Many people are using a toilet (agree) | No | No | No | No | Yes | Yes | No | Yes |
| How many out of 10 think one should use a toilet to defecate? | No | No | No | No | No | No | Yes | Yes |
| All other controls | Yes | Yes | Yes | Yes | Yes | Yes | Yes | Yes |
| Baseline mean at the ward-level | Yes | Yes | Yes | Yes | Yes | Yes | Yes | Yes |
| Observations | 2,562 | 2,562 | 2,562 | 2,562 | 2,562 | 2,562 | 2,562 | 2,562 |

Note: The marginal effects from OLS regression model are presented. The WTP is divided by 1000 and then used in the regression. The standard errors clustered at the ward level are presented in the parenthesis. All other controls include age of the respondent, years of education, gender, marital status, caste affiliation, household size and occupation along with possession of assets listed in Table 1, usage of LPG as a cooking fuel and using a separate room as kitchen. Town panchayat Fixed Effects (FE) are also included in the model. * denotes significance at the 10%, ** at the 5%, and *** at the 1% levels.

These findings suggest that the intervention has been able to alter the EE and NE, which potentially led to increase in the toilet access and usage. However, it should be noted that there might be other channels too. For example, the intervention might have led households to gain better knowledge about the types of toilets to construct, and the associated cost. With higher knowledge and awareness, toilet construction and its usage might also have increased. A potential future work can be to gauge the effects of awareness or knowledge-based interventions and compare them with interventions focussed only on social expectations.

**Further Analysis**

*Effect on toilet ownership and usage*

If the perceived community prevalence and approval of toilet usage through the intervention led to higher toilet usage and access, the perceived valuation for surrounding with low open defecation or residing in ODF areas should also increase. It is important to note that regardless of whether a household uses the toilet, residing in an ODF environment that provides a 'cleaner' neighborhood can potentially result in higher health benefits thereby ensuring a better sense of well-being. A cleaner and hygienic living environment is a public good enjoyed and potentially valued by all residents because it improves public health while enhancing quality of life and productivity.

Accordingly, we assess the effect of the intervention on improving the value placed on residing in areas with a lower prevalence of OD. Here, we measure this value by capturing the Willingness to Pay (WTP) or an "expression of attitude" by residents to live in areas with varying levels of OD. In the endline survey, we particularly ask the following questions:

*(i) "Suppose there are two areas with similar houses you can rent: one, where there is zero open defecation, and another, where many defecate in the open. If you want to live in the area where no one defecates in the open, how much extra are you willing to pay for rent per month?"*

*(ii) "How much extra money per month will you pay to live in an area, where half of the people defecate in the open compared to an area where almost everyone defecates in the open?"*

We use these questions to form our outcome variables. Please note here that we are gauging the valuation for two scenarios, in the first choosing between an area with high OD and an OD-free area, in the second choosing between an area with 50% OD versus a full OD area. This allows us to assess the impact of the intervention on valuing living in OD-free areas and areas with partial OD. From the data collected, we find that the extra rent an average household is willing to pay is INR. 2014 (~ $23) per month as rent for moving from a high OD to an OD free neighborhood. The average WTP for moving to an area with half OD is INR 1136 (~ $13)

Please note that we have asked questions on WTP to estimate economic value of residing in areas with lower prevalence of OD, which is a non-market good. In absence of the possibility of running experiments or auctions with the larger LENNS intervention, we rely on stated preference method through open-ended questions. Apart from eliciting the WTP for moving from a high OD to an OD free area, we also gather information on the WTP for moving to an area with half prevalence of OD as mentioned to ensure that the respondents were able to comprehend the question. Further, a comparison of the WTPs in both these cases can allow to assess if the responses are consistent. For over 87.3% of the respondents, we find that reported WTP for moving from a high OD to ODF neighborhood is greater than that for moving to an area with half OD prevalence. The mean WTP in the former case is INR. 2014 while for the later, it is INR 1136, which indicates that the stated WTP is close to 44% lower. In terms of comprehension, this piece evidence reveals that the respondents were able to understand the questions posed. Nevertheless, in appendix W, we have done multiple checks to ensure that the reported WTPs are valid and consistent.

To estimate the impact on WTP, we use equation 1 but drop $Mean\left(Y_{iwpt=0}\right)$ from the regression as we did not collect information on WTP in the baseline. In the regression, the two measures of WTP as elucidated above is taken as the outcome variables.[13] The findings (Table 5) indicate a statistically significant increase in WTP for residing in ODF communities among respondents from the treated wards. Specifically, we find an average increase in WTP of about Rs. 336 per month ($47

[13] We have divided the reported by 1000 and then presented the estimates for brevity.

annually) for moving from an area with a high prevalence of OD to an area with no prevalence, which can be causally linked to our intervention. We also find that individuals from the intervention wards are willing to pay more for moving to areas with 50% OD from areas with almost 100% open defecation. In this case, the increase in WTP is lower at around Rs. 237 ($34 annually).[14] The estimates using the covariates listed in the pre-analysis plan also remain consistent (Appendix Table A8).

Table 5: Impact on WTP for moving from high OD to ODF and half OD neighborhoods

| | WTP (full OD) | WTP (half OD) | WTP (full OD) | WTP (half OD) |
|---|---|---|---|---|
| | (1) | (2) | (3) | (4) |
| Treatment | 0.334** | 0.240** | 0.316*** | 0.241*** |
| | (0.160) | (0.111) | (0.106) | (0.084) |
| Controls | No | No | Yes | Yes |
| Mean in control areas (endline) | 1.847 | 1.016 | | |
| Observations | 2,571 | 2,571 | 2,562 | 2,562 |
| R-squared | 0.020 | 0.020 | 0.093 | 0.056 |

Note: The marginal effects from LPM model are presented. The WTP is divided by 1000 and then used in the regression. The standard errors clustered at the ward level are presented in the parenthesis. The controls include age of the respondent, years of education, gender, marital status, caste affiliation, household size and occupation along with possession of assets listed in Table 1, usage of LPG as a cooking fuel and using a separate room as kitchen. Town panchayat fixed effects (FE) are also included in the model. * denotes significance at the 10%, ** at the 5%, and *** at the 1% levels.

The cumulative distribution plots of the WTP reported in the treated and control wards indicate that the one for control remains above that for the treated for most part of the distribution (Appendix Figure F1 and Appendix Figure F2). For example, 7 percent in the treated wards reported a WTP (for moving to an area with no OD) of INR 500 or lesser as against 9 percent in the control wards. The share of respondents reporting a WTP of INR 2000 or above in the intervened wards is 41 percent while that in the control ones is 27 percent. Even the share of respondents who are unwilling to pay any money is 3 percent in control wards compared to 2 percent in the treated.

It must be noted that we also collected additional indicators to elicit the WTP for residing in ODF surroundings. Here, we use a discrete choice experiment and start off asking the following question: *"Would you be willing to pay Rs. 2000 extra to live in an area with zero open defecation*

[14] Discussions with the intervention implementing partner indicate that the average rent in the area of our study was around Rs. Rs. 2000 per month during the time of the survey. So, broadly speaking the average WTP to stay in an area with lower prevalence of OD is about 12%-17% more than the average rent paid by the dwellers.

*compared to moving to an area where most people defecate in the open for a similar house?".* If the respondent says "yes" to the above, we further provide the following choices one by one: *Rs. (2200; 2500; 2700; 3000; 3500; 4000; 4500; 5000 and above Rs. 5000)* and stop wherever the respondent says "no". If the response to the above question is "no", we start providing the following choices one by one till the respondent says "yes": *Rs. (1800; 1500; 1200; 1000; 800; 500; 300; 100; below Rs. 100; I would not pay anything).*[15]

We use these set of questions pertaining to the discrete choice experiments to additionally examine whether the intervention has been successful in raising the WTP at least by Rs. 2000. Here, the outcome variable assumes the value of "1" if the reported WTP is Rs. 2000 or above and "0" otherwise; therefore, it is binary in nature. Next, we repeat the same exercise and create additional outcome variables by taking the choice threshold from Rs. 500 to Rs. 5000 instead of Rs. 2000. In other words, this variable takes the value of "1" if the final amount is the threshold amount or above and "0" otherwise. This allows us to assess the intervention effects on increasing the probability of paying a higher amount by the revealed threshold as we increase it. The marginal effects from the regression, along with the 90% confidence interval are shown in figure 1.[16] The findings reveal an inverted U-shaped effect of the intervention (Figure 2). We do not observe any significant difference at the lower end of WTP between the treatment and control arms. For example, there is no significant difference in the likelihood of paying Rs. 500/ Rs. 800 or more between the two groups. However, as the threshold value increases, we observe a disproportionate and significant increase in the likelihood of willingness to pay the amount in the treated wards. Nevertheless, as this value increases, we find that the difference in the likelihood between the two groups starts decreasing and becomes statistically insignificant at the value of Rs. 5000 or more. This is intuitive as it suggests that the intervention has been successful in increasing how people value ODF surroundings, but not beyond a certain level.

Figure 2: Results from the discrete choice experiment

---

[15] In Appendix W, we have also compared the WTP obtained from the open-ended questions with that elicited from the discrete choice experiments. The WTP values show high degree of similarity.

[16] We use the regression as shown in equation 1 but without $Mean(Y_{iwpt=0})$ as a covariate and estimate a Linear Probability Model (LPM) to assess the impact.

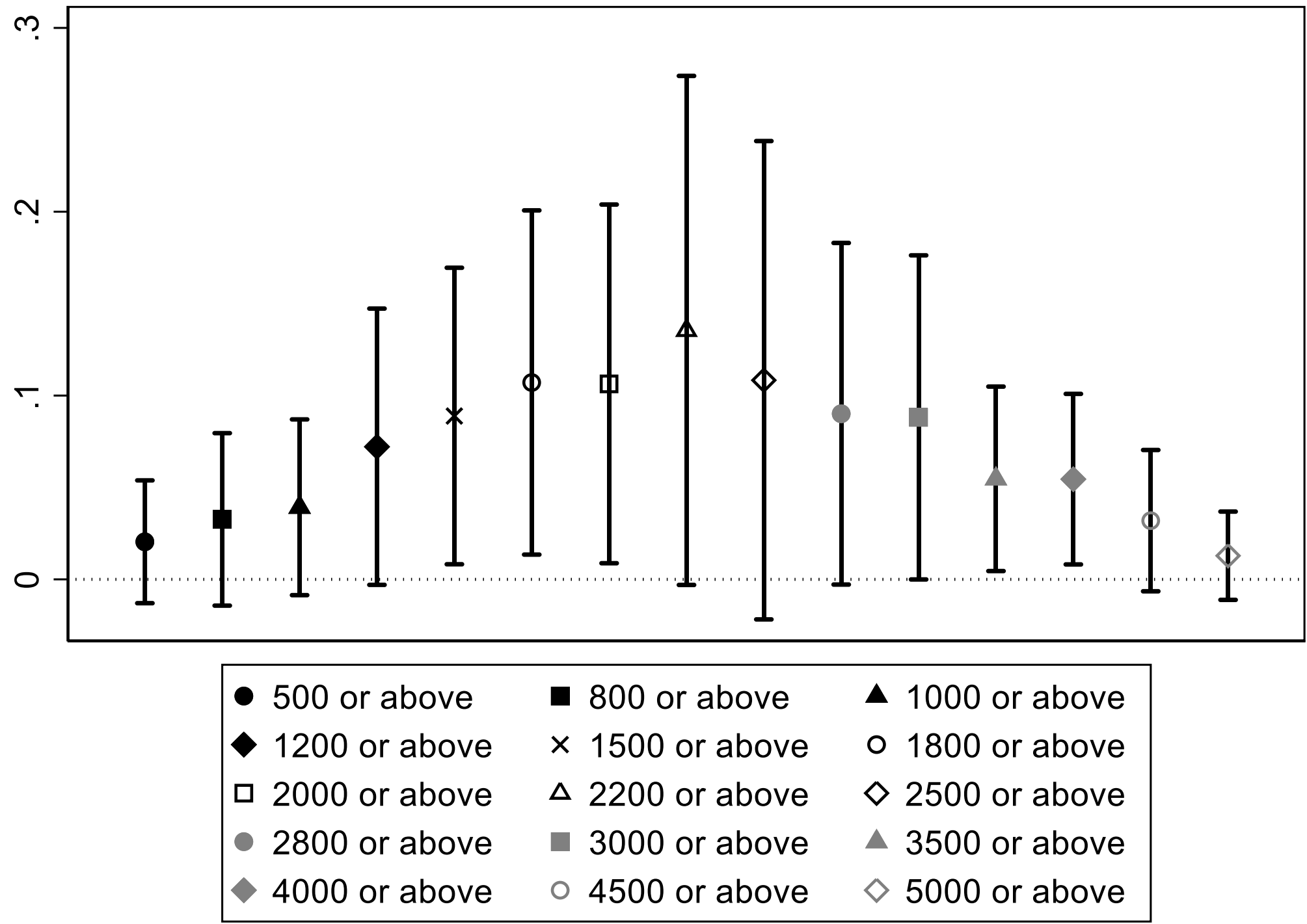

Note: The marginal effects from LPM model are plotted. 95% Confidence Interval calculated by the standard errors, clustered at the ward level are also plotted. The controls include age of the respondent, years of education, gender, marital status, caste affiliation, household size and occupation along with possession of assets listed in Table 1, usage of LPG as a cooking fuel and using a separate room as kitchen. Town panchayat fixed effects are also included in the model.

*Contamination*

One crucial aspect of our intervention involves disseminating information through posters, wall paintings, and household stickers. Consequently, it is likely that individuals from the control wards who visited the treated wards would have been exposed to these materials. Furthermore, there is a possibility that family members or friends residing in the treated wards may have shared information about the intervention, leading to contamination in the control areas. Although we have implemented measures to prevent information spillover by creating a buffer space between the treated and control wards, it may not be entirely effective in ensuring complete isolation. Our data reveals that 66% of respondents from the treated wards are aware of the NN intervention. However, 26% of respondents from the control wards also have heard about the intervention. Consequently, this contamination may introduce a potential bias in the impact estimates of the intervention.

To overcome this, we estimate the LATE to estimate the impact of the actually reported treatment status as reported during the survey on WTP. Because the actual treatment status might be endogenous to the outcome variables, we use the original randomized treatment status, which is exogenous as the IV. Here, we estimate a two-stage least square (2SLS) regression (Angrist and Pishcke, 2008). We use two indicators asked to every respondent, irrespective of whether they reside in the treatment wards to measure the extent to which they were exposed to the intervention. The first one is whether the respondents have heard of *Nam Nalavazhvu* and the second being if they have seen the posters of the intervention. This exposure variable takes the value of 1 if the respondent answers positively for both these questions, 0 for otherwise. The estimates from the 2SLS regressions are given in Table 6. We observe that the likelihood of using toilets increases by 16-17 percentage points for respondents who are exposed to the intervention. Therefore, the estimates without adjusting for the potential contamination are possibly an under-estimate of the treatment effect.

Table 6: LATE estimates to account for contamination

| | Last usage: toilet | Last two days primary defecation place: toilet | Exclusive toilets usage in last two days | Exclusive toilets usage in last seven days | All members have access to toilet |
|---|---|---|---|---|---|
| Intervention exposure | 0.161** | 0.166** | 0.159** | 0.169** | 0.202* |
| | (0.075) | (0.075) | (0.075) | (0.074) | (0.111) |
| All controls | Yes | Yes | Yes | Yes | Yes |
| Baseline mean at the ward-level | Yes | Yes | Yes | Yes | Yes |
| Observations | 2,561 | 2,562 | 2,562 | 2,562 | 2,562 |
| Kleibergen-Paap Wald rk F statistic | 59.88 | 59.67 | 58.69 | 58.17 | 61.54 |

Note: The marginal effects from LPM model are presented. The WTP is divided by 1000 and then used in the regression. The standard errors clustered at the ward level are presented in the parenthesis. *ivreg2* command in STATA is used to estimate the IV regression. The controls include age of the respondent, years of education, gender, marital status, caste affiliation, household size and occupation along with possession of assets such as TV, computer, internet connection, usage of LPG as a cooking fuel and using a separate room as kitchen. Town panchayat Fixed Effects (FE) are also included in the model. * denotes significance at the 10%, ** at the 5%, and *** at the 1% levels.

*Robustness Checks*

Since we were not able to gather information in the endline from the same households surveyed during the baseline, we acknowledge the possibility of a potential bias in our causal estimates. Accordingly, we conducted a number of robustness checks to ensure that the estimates are indeed picking up the effects of the intervention. First, we conduct bounds analysis on the estimated coefficients as suggested by Altonji et al. (2005) and Oster (2019). Second, we make use of the treatment effects estimators with Inverse Probability Weighted (IPW) regression, which can control for the potential selection bias at the treatment stage (Wooldridge 2007). Third, we run all the regressions including only the controls outlined in the pre-analysis plan. Fourth, we excluded the wards affected by the CAA-NRC protest from our analysis. Overall, we find all the regression results to be robust to these checks. The details are given in Appendix R.

## 7. Discussion and Conclusion

Despite substantial improvements in sanitation infrastructure, persistent open defecation in India highlights the importance of behavioral and belief-based constraints in influencing toilet usage for defecation. In this paper, we examine whether shifting social expectations can influence sanitation behavior and the valuation of ODF environments. Using a randomized intervention in peri-urban Tamil Nadu, we provide evidence that changes in beliefs about community behavior can generate significant improvements in both toilet access and consistent usage.

We document three main findings. First, the intervention leads to a significant increase in toilet access and sustained usage. More specifically, we find an increase by 8 and 6-7 percentage points over the control mean value of about 45 percent and 70 percent respectively. Second, the intervention is able to shift the social expectations, particularly EE through higher perceived prevalence of toilet usage within the community and NE through higher perceived community approval. Both of these emerge as key mechanisms driving higher latrine usage for defecation purpose. Third, we observe that these shifts in beliefs and behavior translate into a higher willingness to pay for residing in areas with lower open defecation, indicating that individuals revise upward the perceived value of cleaner sanitation environments.

Taken together, these findings highlight the role of belief-driven mechanisms in shaping both health behavior and the demand for healthy surroundings. In settings where sanitation generates externalities, individual incentives depend critically on perceptions of community compliance. The intervention, by altering these perceptions, increases the expected marginal health return to private sanitation investment and reduces coordination failures that may sustain low-compliance equilibria. The observed increase in willingness to pay further suggests that changes in beliefs affect not only behavior but also how individuals value improvements in their health environment. This assumes importance as it may indicate that there might be positive long-term effects as well.

This study contributes to the literature by demonstrating that behavioral interventions targeting social expectations can complement the other policy tools such as subsidies and infrastructure provision. While prior work has largely focused on reducing financial constraints, our findings show that belief-based interventions can shift both adoption decisions and the valuation of sanitation externalities, even in the absence of direct monetary incentives. From a policy perspective, these results underscore the importance of incorporating expectation-based strategies into sanitation programs. Interventions that make community behavior more salient and credible can help move populations toward high-compliance equilibria and enhance the effectiveness of existing infrastructure investments. Importantly, future research can look into whether similar mechanisms can operate in other domains where health outcomes depend on collective behavior, such as vaccination uptake, hygiene practices, and environmental health.

Finally, the study has some limitations. The use of repeated cross-sectional data prevents us from tracking individual-level behavioral persistence over time, and the external validity of the findings may be context-specific. In addition, while our analysis identifies social expectations as an important channel, other mechanisms such as improved information about sanitation technologies may also contribute to the observed effects. Future research could examine the long-term sustainability of belief-driven behavior change and explore how expectation-based interventions interact with financial incentives in determining health outcomes.

**Appendix A1. Description of the intervention**

The LENNS intervention was deployed in 38 intervention wards in two disticts, Pudukkottai and Karur, across ten town panchayats. The intervention operated at four levels within each intervention cluster: the ward, group, household, and individual levels. A total of 7644 households were engaged for the intervention.

The content or types of information being disseminated changed over time in order to maximize dynamic descriptive messages regarding improved sanitation practices in one's community (Cialdini et al., 2006). Local members were hired as paid outreach workers. They underwent a week of training including in house and community based practice rounds. The socio-behavioral communication materials all highlighted descriptive norms of improved toilet usage, that this ward was changing, aiming to generate a call to action.

*At the ward level:*

**Community Mobilization and Commitment Events (n=55)**

These comprised of community meetings that included men and women of all ages, and also influential political and community leaders to applaud positive deviants and promote improved sanitation practices. Messages highlighted the change happening in their communities and the benefits of using a toilet for families. These reinforced the messages they hear through other group (peer counselling sessions) and household activities. The COVID-19 pandemic limited the number of events deployed due to local congregation limits.

**Mass media broadcasting descriptive information regarding actual and/or prospective adoption of improved sanitation practices via roving announcements (n=380)**

The implementation team deployed ten rounds of audio announcements in each of the 38 wards. The audio content were informed by real-time data regarding people's actual and committed sanitation practices. They also invited community members to peer group session to discuss toilet and community mobilization events.

*At the group level:*

**Peer counselling sessions (n=739)**

These sessions facilitated conversations and information sharing amongst households regarding how to access sanitation markets, different technologies available in the market, barrier identification, planning, and coping by discussing personal experiences, challenges, solutions regarding targeted sanitation practices. These were conducted in small groups at the neighborhood level, and used workbooks with norm focused case studies to engage community members towards solutions.

*At the household level:*

**Household counselling visits (in 7644 households)** allowed outreach workers to inform them about the household's potential for change. The outreach workers used **flipbooks** which contain information on financial resources and toilet designs to enable those who become interested in constructing a toilet. The outreach worker offered to help them achieve their goals by mapping out the next steps. These households were also informed of their neighbors' improved practices. For households with toilets, outreach workers used flipbooks to motivate them to maintain their toilet use behaviors, keep their toilets clean and encourage their neighbors to do the same. All households got sanitation themed board games to encourage children to get involved.

**Visual signals of improved behaviors (n=5260):** Improved behaviors of neighbors were signaled using stickers (one for each promoted behavior) to create awareness of similar other's improved practices. The size and content of these stickers (thumbs up signs) were revised following feedback from the recipient households and their neighbors.

For further details, please refer Ashraf et al. (2021)

The ward outreach workers who are residents of the ward and have at least 12 years of formal education are involved in the implementation of the activities. Field supervisors have also been involved to facilitate ward-level activities.

Table A1: Specific details regarding the aim, frequency, and operationalization of each activity

| Activity | Aim | Target audience | Planned frequency of activity | Actual roll out | Tools, tangibles |
|---|---|---|---|---|---|
| Census ward | | | | | |
| Community mobilization and public commitment events (CMCEs) | The aim of this activity is to:<br>· Generate awareness of the project within the ward and mobilize action<br>· Update factual beliefs (e.g., information regarding where and how to connect with sanitation markets, there are affordable toilet technologies) through the form of contextually appropriate content<br>· Establish public commitment (public signaling of future behavior) to adopt improved sanitation practices | All individuals residing in the ward, especially:<br>1. Non-users & inconsistent users (to ensure their commitment)<br>2. Key change agents – e.g., natural and organizational leaders (to leverage their influence) | 2-3 events per ward; one-time activity*<br><br>* 2-3 events will take place per ward to limit the distance needed to travel within the ward to reach the CMCEs. The events will be targeted geographically. Certain groups of people (e.g., SES, castes) tend to reside together within wards, so conducting several events in different locations within the ward will maximize potential reach to all types of people residing in the ward. | 55 events | · LENNS banner (one per intervention ward) depicting key messages (where names or thumb/hand prints will be placed to publicly display commitment)<br>· Master of ceremony guide<br>· Key messages to be disseminated via song and dance<br>· Skit scripts |
| Mass media broadcasting descriptive information about others' sanitation practices via dynamic roving announcements (i.e., bikes/rickshaws mounted with loud speakers) circulating throughout the ward; routes concentrated on areas with low exclusive use, ownership of toilet facilities | The aim of this activity is to:<br>· Provide dynamic descriptive information about others' sanitation practices (actual or prospective practices) via a dynamic roving signal broadcast throughout the ward. This will facilitate wider and more focally directed signaling within the ward, and facilitate reach to individuals who do not travel to areas where the ward-level visual signals are situated or to those who would not see/understand the visuals depicted (e.g., ppl with physical disabilities, visual impairments) | All individuals residing in the ward, especially those in low coverage/use areas | One week before the CMCEs, one day before the CMCEs, and the day of the CMCEs (as a means of mobilizing people). Weekly for 2 months after the CMCEs, fortnightly thereafter (i.e., twice per month for the remaining 10 months of the intervention period) | Commences one week prior to the CMCE | · Data and messages<br>· Rickshaw/bike drivers |
| Group | | | | | |

| Peer counselling sessions | The aim of this activity is to:<br>· Facilitate conversations & info sharing amongst group members regarding how to access sanitation markets, different technologies available in the market, barrier identification, planning, and coping by discussing personal experiences, challenges, solutions regarding targeted sanitation practices<br>· Enhance action knowledge & capacity through group discussions, cross-fertilization visits<br>· Reinforce messaging from ward, network, and household-level intervention activities<br>· Update vicarious experiences, enact verbal persuasion – I did it, you can too! | At least one adult individual per household in the ward | Three peer counselling sessions per series (at least one adult male and one adult female per household to attend the series of three session), each session in the series to be scheduled two to four weeks apart based on the group members' preferences | 739 sessions completed | · LENNS peer counselling flipbook in which curricula for 3 counselling sessions (one session per target behavior) will be included |
| --- | --- | --- | --- | --- | --- |
| Household | | | | | |
| Household counselling visits and transmission of descriptive information | The aim of this activity is to:<br>· Provide personalized counselling to household members to equip them with the knowledge, skills, motivation necessary to adopt improved sanitation practices<br>· Foster goal setting, self-regulation, action capacity & planning, barrier identification and coping strategies to facilitate adoption and maintenance of improved sanitation practices<br>· Reinforce signaling of sanitation practices through an additional mechanism and intervention level so as to capture individuals who may have been missed by other activities<br>· Give households that did not attend the CMCE a chance to indicate commitment | All households within the ward, with priority going to households without toilet facilities | Monthly | Household visits were frequent where each households without toilets got atleast 1 visit per week. Those who were inconsistent users got a visit twice a month, and regular users were visited atleast once a month. | · Household counselling visit flipbook<br>· LENNS household goal and monitoring cards |

| Cleaning schedule goal setting and monitoring | The aim of this activity is to:<br>· Foster goal setting, self-regulation, action capacity & planning, barrier identification and coping strategies to facilitate sanitation facilities O&M | All households in the census ward with toilet facilities | Monthly monitoring | As needed | · LENNS household goal and monitoring cards (one per household in each intervention ward) |
|---|---|---|---|---|---|
| Individual | | | | | |
| Toilet use monitoring (i.e., data collection on individual sanitation practices) and transmission of descriptive information regarding similar others' sanitation practices to those who are lagging behind | The aim of this activity is to:<br>· Provide information on an individual level to those who are lagging behind in order to facilitate the updating of their empirical expectations<br>· Generate data to inform and guide various LENNS intervention activities and facilitate adaptive management<br>· Tack progress and rates of adoption | · Individuals who are lagging behind (i.e., ppl not exclusively using a toilet)<br>· All individuals (toilet use monitoring) | Monthly | Concurrent with household counselling visits | · Tablets<br>· Monitoring survey<br>· |

Appendix Table A2: Definition of the variables used

| **Variable** | **Definition** |
|---|---|
| *Outcome variable* | |
| *Toilet access and usage* | |
| Some people defecate in the open and some people use a toilet. Where did you defecate the last time you needed to? | = 1 if the response is public/ community toilets, household toilet or any other toilet; 0 otherwise |
| During the last two days, where was your primary place of defecation? | = 1 if the response is public/ community toilets, household toilet or any other toilet; 0 otherwise |
| During the last two days, did you only use a toilet for defecation? | = 1 if the response is yes; 0 otherwise |
| During the last seven days, including today, did you use a toilet for defecation when you needed to defecate? | = 1 if the response is every time; 0 otherwise |
| Do members of your household have access to a private, public, or community toilet? | = 1 if all members in the household have access to a toilet; 0 otherwise. |
| *Willingness to Pay (WTP)* | |
| Suppose there are two areas with similar houses you can rent: one, where there is zero open defecation, and another, where many defecate in the open. If you want to live in the area where no one defecates in the open, how much extra are you willing to pay for rent per month?<br><br>(WTP- full OD) | Reported amount |
| How much extra money per month will you pay to live in an area, where half of the people defecate in the open compared to an area where almost everyone defecates in the open?<br><br>(WTP- half OD) | Reported amount |
| Would you be willing to pay Rs. 2000 extra to live in an area with zero open defecation compared to moving to an area where most people defecate in the open for a similar house?<br><br>(Discrete choice experiment) | If the respondent says "yes" to the above, we further provide the following choices one by one: Rs. (2200; 2500; 2700; 3000; 3500; 4000; 4500; 5000 and above Rs. 5000) and stop wherever the respondent says "no". If the response to the above question is "no", we start providing the following choices one by one till the respondent says "yes": Rs. (1800; 1500; 1200; 1000; 800; 500; 300; 100; below Rs. 100; I would not pay anything).<br><br>== 1 if the reported WTP is Rs. 2000 or above; 0 otherwise |
| | = 1 if the reported WTP is Rs. X; 0 otherwise (here X varies from 500, 800, 1000, 1200, 1500, 1800, 2200, 2500, 2700, 3000, 3500, 4000, 4500, 5000) |
| *Main variable of interest* | |
| Treatment | =1 if the ward where the surveyed household resides is treated; 0 otherwise |
| *Control variables* | |
| Age | Reported age of the respondent (in years) |
| Sex | = 1 if the respondent is female; 0 otherwise |
| Upper caste | = 1 if the respondent belongs to upper caste; 0 otherwise |
| Household size | Number of members in the household |
| Household asset index | First component generated through Principal Component Analysis (PCA) of possession of the following assets: color TV, mobile phone, motorcycle/scooter, fridge, internet, computer, AC/cooler, and washing machine. |
| Additional controls | |
| Education | Reported years of schooling |
| Marital status | =1 if the respondent is currently married; 0 otherwise |
| Owns color TV | = 1 if the respondent has a color TV in his/ her house; 0 otherwise |
| Has internet connection | = 1 if the respondent has access to internet connection; 0 otherwise |
| Owns computer | = 1 if the respondent has a computer or laptop in his/ her house; 0 otherwise |
| Owns air-conditioner | = 1 if the respondent has an air-conditioner/ cooler in his/ her house; 0 otherwise |
| Owns washing machine | = 1 if the respondent has a washing machine in his/ her house; 0 otherwise |
| Owns mobile | = 1 if the respondent's household owns at least one functioning mobile phone; 0 otherwise |
| Owns motorcycle | = 1 if the respondent has a motorcycle in the household; 0 otherwise |
| Owns fridge | = 1 if the respondent has a fridge or refrigerator in the household; 0 otherwise |
| Owns car | = 1 if the respondent has a car in the household; 0 otherwise |
| Owns dish antenna | = 1 if there is a dish antenna fixed in the household; 0 otherwise |
| Gas as cooking fuel | = 1 if the household of the respondent uses Liquefied Petroleum Gas (LPG) as the main cooking fuel; 0 otherwise |
| Has separate kitchen | = 1 if the household of the respondent has a separate room for kitchen; 0 otherwise |
| Occupation: laborer | = 1 if the principal occupation of the respondent is labor works; 0 otherwise |
| PCA asset index (Pre-Analysis Plan) | Principal Component Analysis (PCA) with eth first component using the following indicators about ownership of assets: Owns color TV; Owns mobile; Owns motorcycle ; Owns fridge; Has internet connection; Owns computer; Owns air-conditioner; Owns washing machine (all defined above) |
| *Social Expectations* | |
| Out of ten members in your ward, how many do you think use a toilet every time they need to defecate? | Reported number |
| Out of ten households in your ward, how many do you think own a toilet at home? | Reported number |
| Statement:<br><br>Many people in my ward built new toilets in the past six months | = 1 if the respondent agrees with this statement; 0 otherwise |
| Statement:<br><br>I think more people use a toilet in my ward compared to six months ago | = 1 if the respondent agrees with this statement; 0 otherwise |
| Out of ten members of your ward, how many do you think believe one should use a toilet to defecate? | Reported number |

Appendix Table A3: Impact on toilet use and access

| | Last usage: toilet | Last two days primary defecation place: toilet | Exclusive toilets usage in last two days | Exclusive toilets usage in last seven days | All members have access to toilet | Last usage: toilet | Last two days primary defecation place: toilet | Exclusive toilets usage in last two days | Exclusive toilets usage in last seven days | All members have access to toilet |
|---|---|---|---|---|---|---|---|---|---|---|
| | (1) | (2) | (3) | (4) | (5) | (6) | (7) | (8) | (9) | (10) |
| Treatment | 0.104** | 0.105** | 0.104** | 0.113** | 0.124** | 0.064** | 0.067** | 0.064** | 0.067** | 0.081* |
| | (0.048) | (0.048) | (0.048) | (0.049) | (0.054) | (0.029) | (0.029) | (0.029) | (0.028) | (0.042) |
| Age | | | | | | 0.003*** | 0.003*** | 0.003*** | 0.003*** | 0.001 |
| | | | | | | (0.001) | (0.001) | (0.001) | (0.001) | (0.001) |
| Years of education | | | | | | 0.010*** | 0.010*** | 0.010*** | 0.010*** | 0.007*** |
| | | | | | | (0.002) | (0.002) | (0.002) | (0.002) | (0.003) |
| Gender: Female | | | | | | 0.032* | 0.030* | 0.030* | 0.028 | 0.068** |
| | | | | | | (0.018) | (0.018) | (0.018) | (0.018) | (0.030) |
| Currently married | | | | | | 0.025 | 0.024 | 0.027 | 0.038* | -0.023 |
| | | | | | | (0.019) | (0.020) | (0.020) | (0.020) | (0.024) |
| Upper caste | | | | | | -0.041 | -0.042 | -0.053 | -0.086*** | -0.091** |
| | | | | | | (0.035) | (0.034) | (0.032) | (0.030) | (0.041) |
| Household size | | | | | | -0.011* | -0.009 | -0.011* | -0.007 | 0.006 |
| | | | | | | (0.006) | (0.006) | (0.006) | (0.007) | (0.008) |
| Owns color TV | | | | | | -0.064 | -0.068 | -0.077 | -0.081 | -0.133** |
| | | | | | | (0.053) | (0.052) | (0.053) | (0.059) | (0.056) |
| Has internet connection | | | | | | 0.036* | 0.036* | 0.034* | 0.013 | -0.130*** |
| | | | | | | (0.020) | (0.020) | (0.020) | (0.022) | (0.027) |
| Owns computer | | | | | | 0.014 | 0.015 | 0.019 | 0.016 | -0.034 |
| | | | | | | (0.026) | (0.027) | (0.026) | (0.028) | (0.048) |
| Owns air-conditioner | | | | | | -0.018 | -0.016 | -0.024 | -0.028 | 0.066 |
| | | | | | | (0.028) | (0.028) | (0.030) | (0.033) | (0.060) |
| Owns washing machine | | | | | | 0.016 | 0.018 | 0.015 | 0.026 | -0.012 |
| | | | | | | (0.019) | (0.019) | (0.020) | (0.022) | (0.040) |
| Owns mobile phone | | | | | | -0.047 | -0.051 | -0.032 | -0.035 | -0.024 |
| | | | | | | (0.044) | (0.044) | (0.046) | (0.047) | (0.053) |
| Owns motorcycle | | | | | | 0.039 | 0.035 | 0.039 | 0.049* | 0.050* |
| | | | | | | (0.024) | (0.025) | (0.023) | (0.026) | (0.029) |
| Owns refrigerator | | | | | | 0.089*** | 0.085*** | 0.088*** | 0.129*** | 0.176*** |
| | | | | | | (0.017) | (0.018) | (0.019) | (0.025) | (0.025) |
| Owns car | | | | | | 0.023 | 0.022 | 0.035 | 0.050 | 0.013 |
| | | | | | | (0.035) | (0.036) | (0.035) | (0.040) | (0.073) |
| Owns dish antenna | | | | | | 0.026 | 0.026 | 0.027 | 0.029 | -0.009 |
| | | | | | | (0.019) | (0.019) | (0.019) | (0.019) | (0.036) |
| Gas as cooking fuel | | | | | | 0.273*** | 0.273*** | 0.275*** | 0.258*** | 0.151*** |
| | | | | | | (0.043) | (0.043) | (0.042) | (0.043) | (0.040) |
| Has separate kitchen | | | | | | 0.164*** | 0.169*** | 0.164*** | 0.126*** | 0.241*** |
| | | | | | | (0.027) | (0.027) | (0.028) | (0.029) | (0.030) |
| Occupation: laborer | | | | | | -0.011 | -0.021 | -0.017 | -0.045* | -0.022 |
| | | | | | | (0.025) | (0.026) | (0.025) | (0.027) | (0.027) |
| Town Panchayat FE | Yes | Yes | Yes | Yes | Yes | Yes | Yes | Yes | Yes | Yes |
| Constant | 0.697*** | 0.694*** | 0.688*** | 0.641*** | 0.453*** | 0.255** | 0.251** | 0.219* | 0.233* | 0.222 |
| | (0.036) | (0.036) | (0.036) | (0.034) | (0.035) | (0.123) | (0.115) | (0.119) | (0.126) | (0.163) |
| Observations | 2,570 | 2,571 | 2,571 | 2,571 | 2,571 | 2,561 | 2,562 | 2,562 | 2,562 | 2,562 |
| R-squared | 0.014 | 0.015 | 0.014 | 0.015 | 0.015 | 0.303 | 0.303 | 0.296 | 0.294 | 0.223 |

Note: The marginal effects from LPM model are presented. The WTP is divided by 1000 and then used in the regression. The standard errors clustered at the ward level are presented in the parenthesis. The covariates from pre-analysis plan include age of the respondent, gender, caste affiliation and household size along with the first PCA component of possession of the following assets: color TV, mobile phone, motorcycle/scooter, fridge, internet, computer, AC/cooler, and washing machine. * denotes significance at the 10%, ** at the 5%, and *** at the 1% levels.

Appendix Figure F1: Cumulative distribution plots for WTP (no OD)

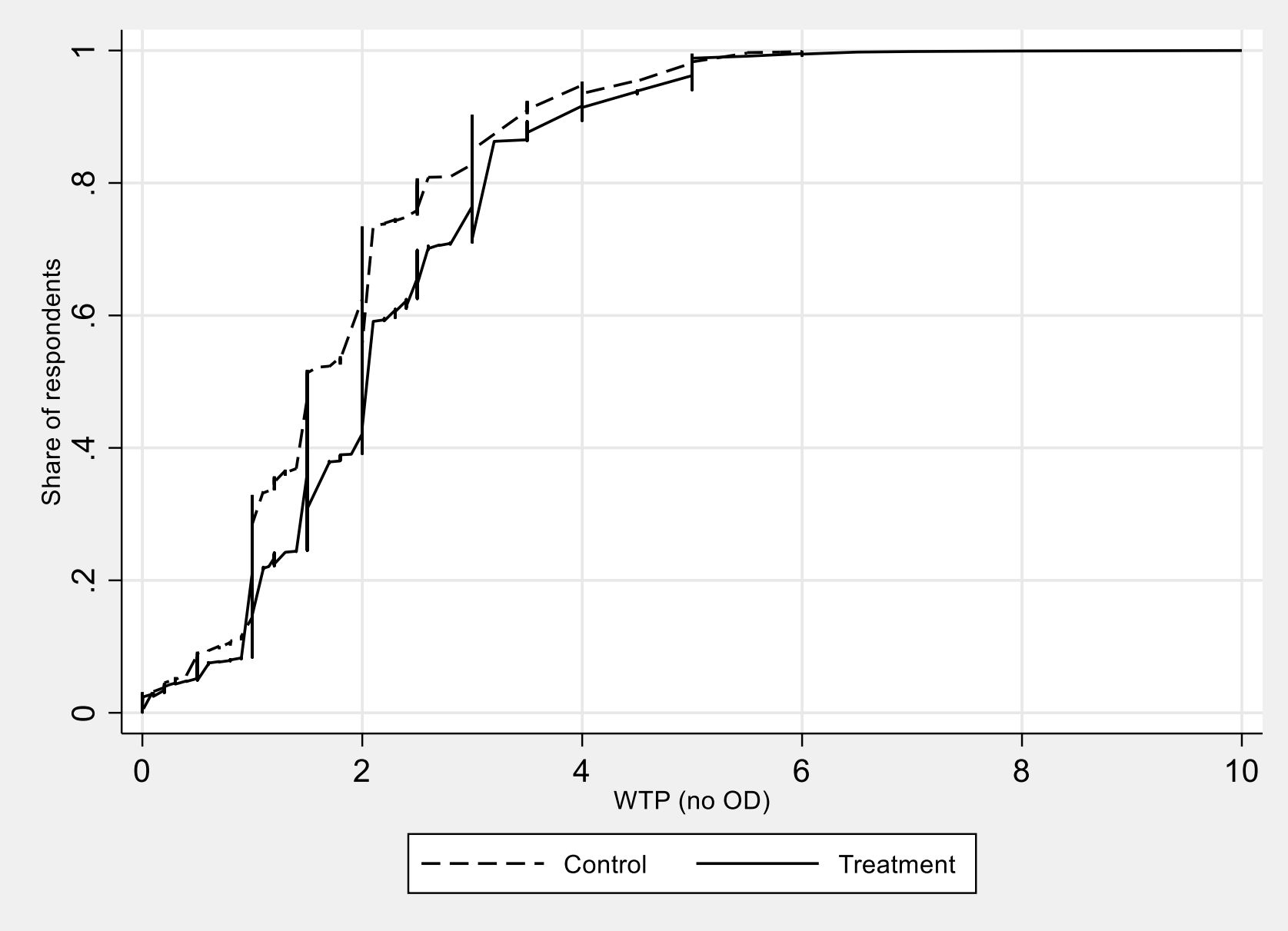

Note. The WTP is divided by 1000 and then used in the regression.

Appendix Figure F2: Cumulative distribution plots for WTP (half OD)

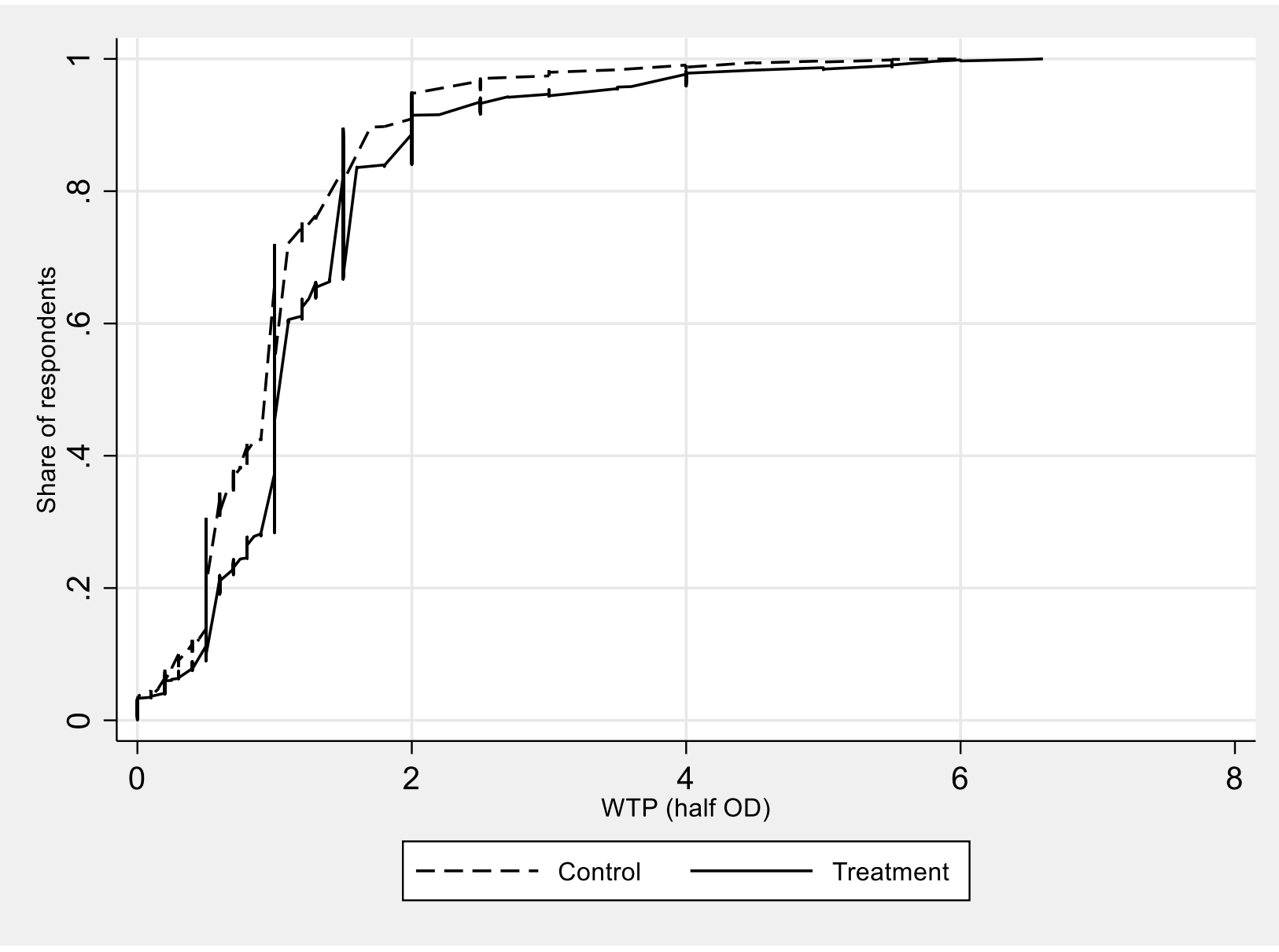

Note. The WTP is divided by 1000 and then used in the regression.

Appendix W: Willingness to pay

We first use open ended questions to elicit the Willingness to Pay (WTP) for moving from a neighborhood with high prevalence of OD to an area with lower prevalence. As indicated in the main manuscript, the questions asked to the respondents are as follows:
*(i) "Suppose there are two areas with similar houses you can rent: one, where there is zero open defecation, and another, where many defecate in the open. If you want to live in the area where no one defecates in the open, how much extra are you willing to pay for rent per month?"*

*(ii) "How much extra money per month will you pay to live in an area, where half of the people defecate in the open compared to an area where almost everyone defecates in the open?"*

The following table presents the descriptive statistics of the full sample of respondents and then separated by those from the treated and control wards (Appendix table A5). Substantially higher extra WTP for residing in ODF areas is observed. We find a fall in the WTP for residing in an area with half OD when compared to residing in areas with no OD by INR. 878 on average. This equals to a decline of about 44%. For treatment areas, the former is about INR 925 lower than the latter and for control wards, it is lower by INR. 831.

Appendix Table A4: Descriptive statistics from open ended questions

| | Full sample | | Treatment | | Control | |
|---|---|---|---|---|---|---|
| | WTP (full OD) | WTP (half OD) | WTP (full OD) | WTP (half OD) | WTP (full OD) | WTP (half OD) |
| Mean | 2014.42 | 1136.36 | 2180.80 | 1255.84 | 1846.61 | 1015.84 |
| Median | 2000 | 1000 | 2000 | 1000 | 1500 | 1000 |
| Standard deviation | 1195.74 | 847.99 | 1231.3 | 936.06 | 1134.79 | 729.56 |

Next, we look at the relationship between the WTP for residing in an area with no OD and that with half OD. The scatter plot is given below in Appendix Figure F4. The linear fit line indicates that the relationship is positive with pairwise correlation of 0.6. This is also statistically significant at 1% level.

Appendix Figure F3: Scatter plot

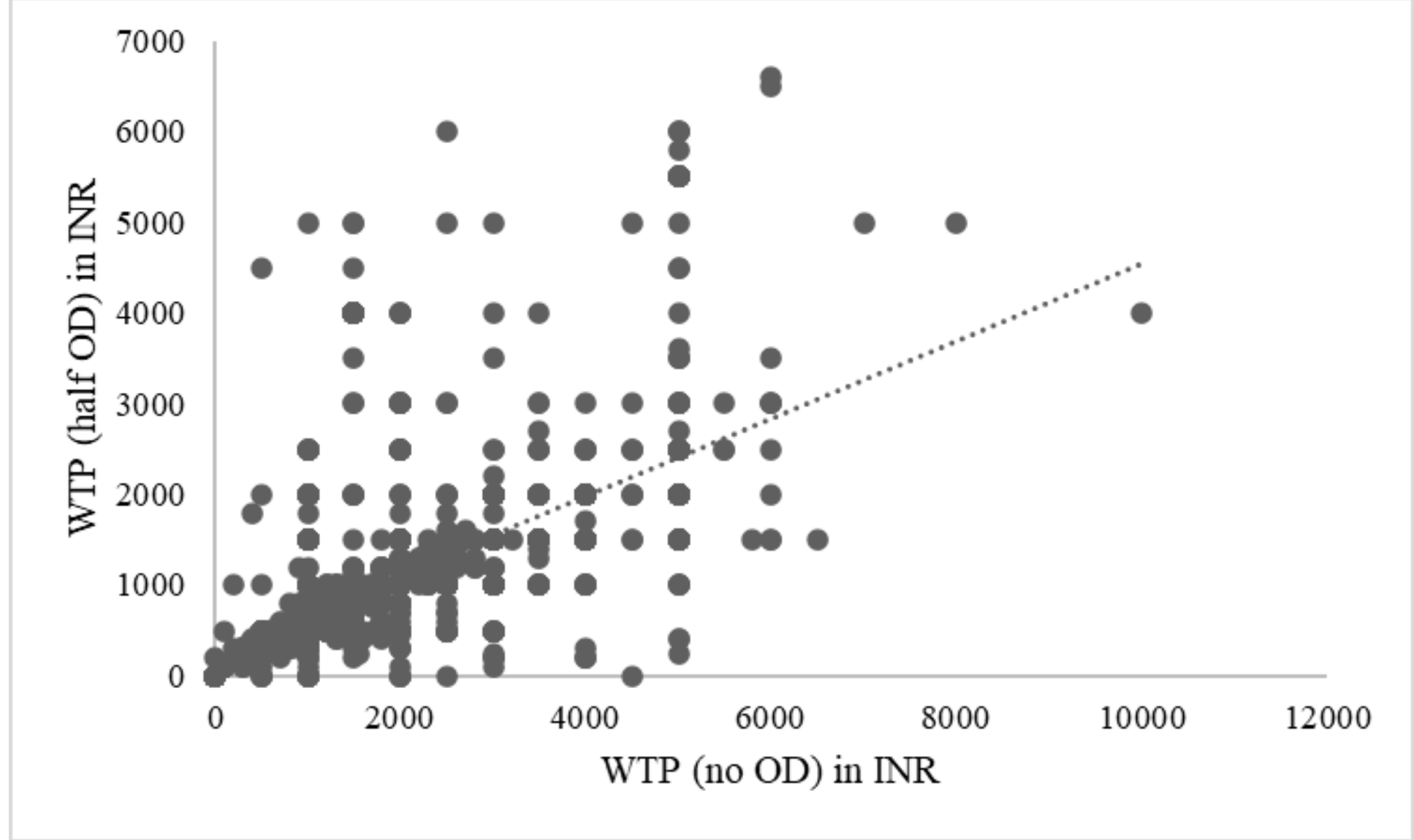

Further, we look at the correlates of the WTP and explore if personal characteristics which should intuitively be associated are indeed observable from the data. The associations are found to be on expected lines. Economic characteristics like owning color television or using internet along with having separate kitchen or using clean fuel for cooking are all positively correlated with WTP and the relationship is statistically significant. Education also is found to have a positive and significant relationship. Further, female respondents report significantly higher WTP, which is intuitive as studies show higher marginal gains from having access to toilets (Das et al. 2023a).

As indicated in the manuscript, we gathered additional data to elicit the WTP for residing in ODF surroundings. We conduct a discrete choice experiment and asked the following question: *"Would you be willing to pay Rs. 2000 extra to live in an area with zero open defecation compared to moving to an area where most people defecate in the open for a similar house?".* If the respondent says "yes" to the above, we further provide the following choices one by one: *Rs. (2200; 2500; 2700; 3000; 3500; 4000; 4500; 5000 and above Rs. 5000)* and stop wherever the respondent says "no". If the response to the above question is "no", we start providing the following choices one by one till the respondent says "yes": *Rs. (1800; 1500; 1200; 1000; 800; 500; 300; 100; below Rs. 100; I would not pay anything).*

From the data, we generate the following descriptive statistics based on the amount where the respondent says "no". If any respondent says “yes” to “above Rs. 5000”, we take the reported WTP to be Rs. 5500. The following table (Appendix table A6) presents these statistics for the full sample of respondents and then separated by those from the treated and control wards.

Appendix Table A5: Descriptive statistics from discrete choice questions

| | Full sample | Treatment wards | Control wards |
|---|---|---|---|
| Mean | 2145.91 | 2304.76 | 1985.70 |
| Median | 2200 | 2200 | 2000 |
| Standard deviation | 1080.46 | 1101.27 | 1035.05 |

It is important to note that despite using an alternative method of eliciting the WTP, this measure is highly correlated with the WTP that we are able to elicit from the open-ended question on moving to area with no open defecation. This is clear from the scatter-plot below. The linear fit regression line indicates a positive relation, which is significant at 1% level.

Appendix Figure F4: Scatter plot

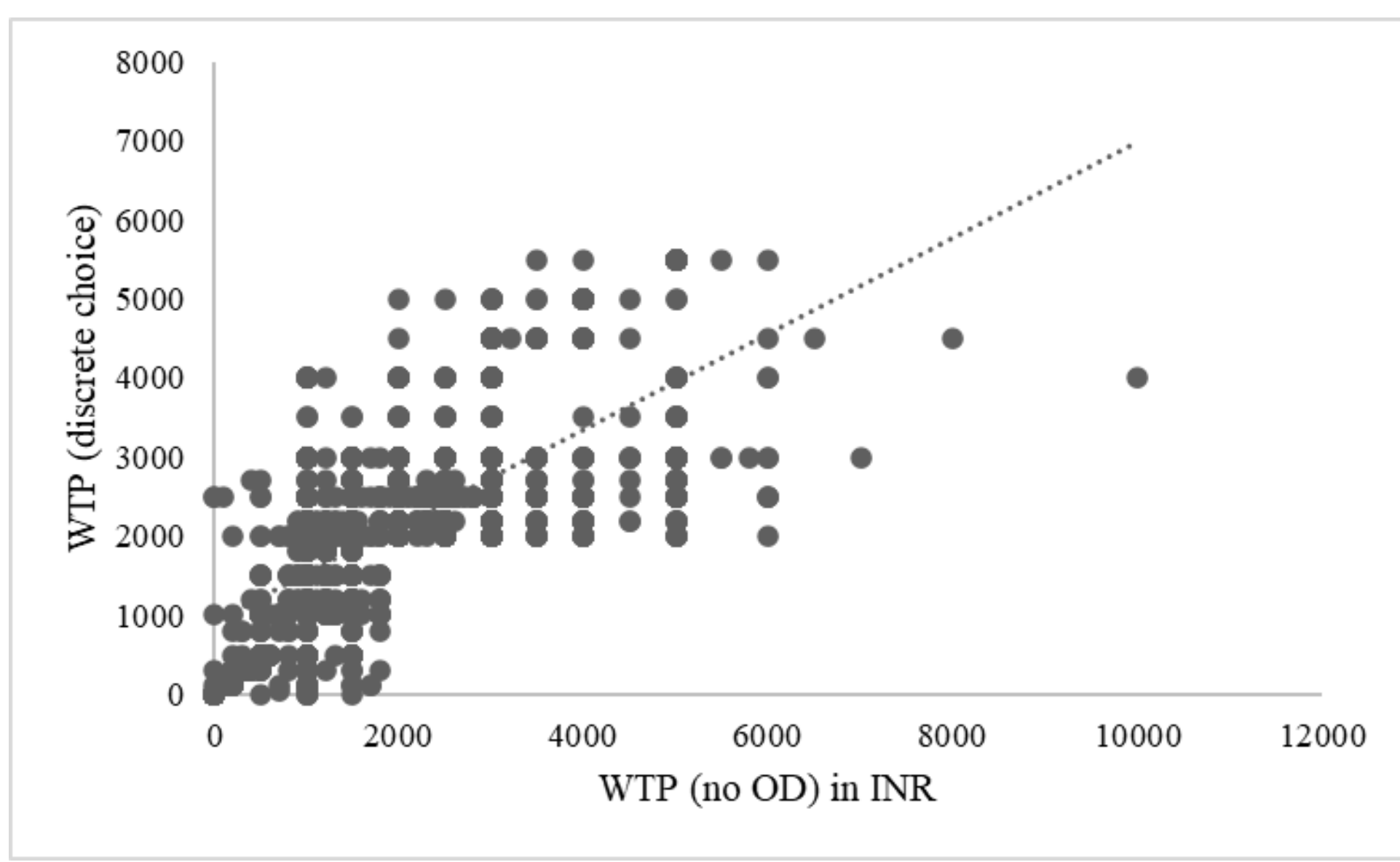

Appendix Table A6a: Correlations with WTP for moving from high OD to ODF neighborhoods

| | (1) | (2) | (3) | (4) | (5) | (6) | (7) | (8) | (9) |
|---|---|---|---|---|---|---|---|---|---|
| Age | -0.004** | | | | | | | | |
| | (0.002) | | | | | | | | |
| Years of education | | 0.027*** | | | | | | | |
| | | (0.004) | | | | | | | |
| Gender: female | | | 0.106** | | | | | | |
| | | | (0.047) | | | | | | |
| Currently married | | | | 0.030 | | | | | |
| | | | | (0.056) | | | | | |
| Upper caste | | | | | -0.051 | | | | |
| | | | | | (0.051) | | | | |
| Household size | | | | | | 0.118*** | | | |
| | | | | | | (0.017) | | | |
| Gas as cooking fuel | | | | | | | 0.129** | | |
| | | | | | | | (0.064) | | |
| Has separate kitchen | | | | | | | | 0.261*** | |
| | | | | | | | | (0.050) | |
| Occupation: laborer | | | | | | | | | -0.170*** |
| | | | | | | | | | (0.062) |
| Constant | 2.192*** | 1.805*** | 1.958*** | 1.991*** | 2.027*** | 1.579*** | 1.900*** | 1.809*** | 2.045*** |
| | (0.078) | (0.039) | (0.035) | (0.049) | (0.028) | (0.066) | (0.058) | (0.042) | (0.026) |
| Observations | 2,571 | 2,571 | 2,571 | 2,568 | 2,565 | 2,571 | 2,571 | 2,571 | 2,571 |
| R-squared | 0.002 | 0.014 | 0.002 | 0.000 | 0.000 | 0.019 | 0.001 | 0.008 | 0.003 |

*Note. Coefficients from OLS regressions are presented along with robust standard errors in parentheses. . * denotes significance at the 10%, ** at the 5%, and *** at the 1% levels.*

Appendix Table A6b: Correlations with WTP for moving from high OD to ODF neighborhoods

| | (1) | (2) | (3) | (4) | (5) | (6) | (7) | (8) | (9) | (10) | (11) |
|---|---|---|---|---|---|---|---|---|---|---|---|
| Owns color TV | 0.230** | | | | | | | | | | |
| | (0.111) | | | | | | | | | | |
| Has internet connection | | 0.365*** | | | | | | | | | |
| | | (0.047) | | | | | | | | | |
| Owns computer | | | 0.122 | | | | | | | | |
| | | | (0.124) | | | | | | | | |
| Owns air-conditioner | | | | 0.516*** | | | | | | | |
| | | | | (0.141) | | | | | | | |
| Owns washing machine | | | | | 0.317*** | | | | | | |
| | | | | | (0.093) | | | | | | |
| Owns mobile phone | | | | | | 0.364*** | | | | | |
| | | | | | | (0.103) | | | | | |
| Owns motorcycle | | | | | | | 0.382*** | | | | |
| | | | | | | | (0.048) | | | | |
| Owns refrigerator | | | | | | | | 0.498*** | | | |
| | | | | | | | | (0.048) | | | |
| Owns car | | | | | | | | | 0.735*** | | |
| | | | | | | | | | (0.186) | | |
| Owns dish antenna | | | | | | | | | | 0.020 | |
| | | | | | | | | | | (0.049) | |
| PCA (asset index) | | | | | | | | | | | 0.167*** |
| | | | | | | | | | | | (0.018) |
| Constant | 1.793*** | 1.834*** | 2.006*** | 1.988*** | 1.980*** | 1.665*** | 1.728*** | 1.790*** | 1.994*** | 2.007*** | 2.006*** |
| | (0.108) | (0.029) | (0.024) | (0.024) | (0.024) | (0.100) | (0.038) | (0.026) | (0.024) | (0.030) | (0.023) |
| Observations | 2,571 | 2,571 | 2,571 | 2,571 | 2,571 | 2,571 | 2,571 | 2,571 | 2,571 | 2,571 | 2,571 |
| R-squared | 0.001 | 0.023 | 0.001 | 0.009 | 0.007 | 0.003 | 0.019 | 0.043 | 0.010 | 0.000 | 0.042 |

*Note. Coefficients from OLS regressions are presented along with robust standard errors in parentheses. . * denotes significance at the 10%, ** at the 5%, and *** at the 1% levels.*

Appendix Table A7a: Correlations with WTP for moving from high OD to half OD neighborhoods

| | (1) | (2) | (3) | (4) | (5) | (6) | (7) | (8) | (9) |
|---|---|---|---|---|---|---|---|---|---|
| Age | -0.002 | | | | | | | | |
| | (0.001) | | | | | | | | |
| Years of education | | 0.005* | | | | | | | |
| | | (0.003) | | | | | | | |
| Gender: female | | | 0.103*** | | | | | | |
| | | | (0.033) | | | | | | |
| Currently married | | | | 0.050 | | | | | |
| | | | | (0.037) | | | | | |
| Upper caste | | | | | -0.055 | | | | |
| | | | | | (0.035) | | | | |
| Household size | | | | | | 0.026** | | | |
| | | | | | | (0.011) | | | |
| Gas as cooking fuel | | | | | | | 0.199*** | | |
| | | | | | | | (0.036) | | |
| Has separate kitchen | | | | | | | | 0.218*** | |
| | | | | | | | | (0.034) | |
| Occupation: laborer | | | | | | | | | 0.039 |
| | | | | | | | | | (0.050) |
| Constant | 1.222*** | 1.094*** | 1.082*** | 1.097*** | 1.150*** | 1.039*** | 0.961*** | 0.965*** | 1.129*** |
| | (0.055) | (0.029) | (0.024) | (0.031) | (0.020) | (0.045) | (0.031) | (0.027) | (0.018) |
| Observations | 2,571 | 2,571 | 2,571 | 2,568 | 2,565 | 2,571 | 2,571 | 2,571 | 2,571 |
| R-squared | 0.001 | 0.001 | 0.004 | 0.001 | 0.001 | 0.002 | 0.006 | 0.011 | 0.000 |

*Note. Coefficients from OLS regressions are presented along with robust standard errors in parentheses. . * denotes significance at the 10%, ** at the 5%, and *** at the 1% levels.*

Appendix Table A7b: Correlations with WTP for moving from high OD to half OD neighborhoods

| | (1) | (2) | (3) | (4) | (5) | (6) | (7) | (8) | (9) | (10) | (11) |
|---|---|---|---|---|---|---|---|---|---|---|---|
| Owns color TV | -0.013 | | | | | | | | | | |
| | (0.084) | | | | | | | | | | |
| Has internet connection | | 0.165*** | | | | | | | | | |
| | | (0.033) | | | | | | | | | |
| Owns computer | | | 0.003 | | | | | | | | |
| | | | (0.084) | | | | | | | | |
| Owns air-conditioner | | | | 0.047 | | | | | | | |
| | | | | (0.084) | | | | | | | |
| Owns washing machine | | | | | 0.026 | | | | | | |
| | | | | | (0.056) | | | | | | |
| Owns mobile phone | | | | | | 0.111 | | | | | |
| | | | | | | (0.073) | | | | | |
| Owns motorcycle | | | | | | | 0.187*** | | | | |
| | | | | | | | (0.033) | | | | |
| Owns refrigerator | | | | | | | | 0.110*** | | | |
| | | | | | | | | (0.034) | | | |
| Owns car | | | | | | | | | 0.164 | | |
| | | | | | | | | | (0.107) | | |
| Owns dish antenna | | | | | | | | | | -0.091*** | |
| | | | | | | | | | | (0.033) | |
| PCA (asset index) | | | | | | | | | | | 0.048*** |
| | | | | | | | | | | | (0.012) |
| Constant | 1.149*** | 1.054*** | 1.136*** | 1.134*** | 1.134*** | 1.029*** | 0.996*** | 1.087*** | 1.132*** | 1.172*** | 1.134*** |
| | (0.082) | (0.021) | (0.017) | (0.017) | (0.018) | (0.071) | (0.026) | (0.022) | (0.017) | (0.023) | (0.017) |
| Observations | 2,571 | 2,571 | 2,571 | 2,571 | 2,571 | 2,571 | 2,571 | 2,571 | 2,571 | 2,571 | 2,571 |
| R-squared | 0.000 | 0.009 | 0.000 | 0.000 | 0.000 | 0.001 | 0.009 | 0.004 | 0.001 | 0.003 | 0.007 |

*Note. Coefficients from OLS regressions are presented along with robust standard errors in parentheses. . * denotes significance at the 10%, ** at the 5%, and *** at the 1% levels.*

**Appendix R: Robustness check**

R1. Bias adjusted treatment effect

To estimate the impact of the intervention, we utilize its random implementation across the wards. The comparison of socio-economic and demographic characteristics along with indicators related to toilet usage across the treated and control wards as shown in table 1 indicate the two groups are well-balanced. Nevertheless, we find that one variable (years of education) is not balanced with the difference significant at a 5% level. The difference in gender and marital status is significant at a 10% level. In absence of the indicators on WTP in the baseline wave, we cannot comment on the balance between the two arms in terms of these indicators. Further, with the onset of the COVID-19 pandemic, we were not able to survey the household surveyed in the baseline and had to depend on repeated cross-sectional data to evaluate the impact. This might also introduce some bias in our estimates.

To address this issue, we make use of the "selection on unobservables versus observable" approach to ensure that the treatment effects account for these potential biases (Altonji et al., 2005; Oster, 2019). The underlying assumption is that the extent of selection on unobservables is related proportionally to that on observables, the ratio of which is given by a parameter δ. Another parameter ($R^2_{max}$), which is the hypothetical explained variation in a regression that includes all possible observed and unobserved characteristics. Oster (2019) simulated a number of existing randomized experiments to propose that the $R^2_{max} = 1.3 * R^2_0$, where, $R^2_0$ indicates the *R*-squared value of the regression model that uses only the observed control variables. Next, with a given $R^2_0$ and $R^2_{max}$, the value of $\delta$ is calculated such that the marginal effect is zero. A value of $|\delta|$, which is greater than 1 implies that the explanatory power of the unobservables is higher than that of the controlled covariates. In a setting like ours, where the intervention has been randomly implemented, the selection bias through the unobservables is likely to be low if not negligible. Because of this, if we find the $|\delta| > 1$, the role of unobservables would have to be very high to turn the intervention effects to be zero. The values of δ, calculated for each of the two main regressions as shown in Appendix Table A9 (columns 1 and 2) are found to be close to 27 and 83, which is much higher than the accepted benchmark of 1. Even for the other outcomes indicators pertaining to EE, NE and toilet usage, the value of $\delta$ is atleast 4. This confirms additionally that our estimates capture the treatment effects and are not confounded by other factors.[1]

R.2 Inverse Probability Weighting and Pre-Analysis Plan

To further account for the potential bias resulting from the issues discussed above, we make use of the treatment effects estimators with Inverse Probability Weighted (IPW) regression, which can control for the potential selection bias at the treatment stage (Wooldridge 2007). Appendix Table A8 presents the estimates from IPW. The findings indicate a significant increase in WTP because of the intervention. We also observe significant increase in EE and NE along with toilet access and usage.

In addition, we run another robustness check using just the covariates that we state in the pre-analysis plan before the endline (https://cdn.clinicaltrials.gov/large-docs/24/NCT04269824/SAP_000.pdf). Accordingly, we construct a household asset index using principal component analysis (PCA) by considering possession of assets such as color TV, mobile phone, motorcycle/scooter, fridge, internet, computer, AC/cooler, and washing machine. The results remain similar qualitatively (Appendix Table A8).

R.3 Excluding wards affected by CAA-NRC protest

Back in December 2019, the Government of India enacted the Citizenship Amendment Act (CAA) Bill, which amends the Indian citizenship act to accept non-Muslim migrants from Afghanistan, Pakistan, and Bangladesh who entered India before 2014 following religious persecutions. The National Register of Citizens (NRC) will officially record every legal citizen of India, which requires a set of prescribed documents.[2] Following these, major protests led by Muslims across the country broke out and continued from December 2019 to March 2020.[3] Because of this, in three out of the seventy-six Muslim-dominated wards, many households did not give consent to be surveyed and we had to replace

[1] This method has been used by a number of studies to examine the possibility of selection on unobservables in different contexts (Alesina et al. 2016; Michalopoulos and Papaioannou, 2016).

[2] More information on CAA can be accessed from https://egazette.nic.in/WriteReadData/2019/214646.pdf. More information on NRC can be obtained from https://blogs.lse.ac.uk/humanrights/2020/08/10/the-national-register-of-citizens-and-indias-commitment-deficit-to-international-law/ (accessed on September 7, 2022).

[3] CAA-NRC protest has been covered across news channels and newspapers. For example https://indianexpress.com/about/caa-protest/ (accessed on September 7, 2022).

them with Hindu households within the same ward during the baseline wave. This compromised the random selection of households for the survey in these three wards. Nevertheless, even if we drop these three wards and re-run the main regressions, our inference on the intervention effects holds (Appendix Table A8)

Appendix Table A8: Bias-adjusted treatment effects, Inverse Probability Weighting, Pre-Analysis Plan and results exclusing wards affected by CAA-NRC protest

| | Last usage: toilet | Last two days primary defecation place: toilet | Exclusive toilets usage in last two days | Exclusive toilets usage in last seven days | All members have access to toilet | WTP (full OD) | WTP (half OD) | Many people are constructing a toilet | How many out of 10 own a toilet? | Many people are using a toilet | How many out of 10 think one should use a toilet to defecate? |
|---|---|---|---|---|---|---|---|---|---|---|---|
| Bias-adjusted treatment effects | | | | | | | | | | | |
| Delta ($\boldsymbol{\delta}$) | 4.5 | 4.6 | 4.1 | 3.9 | 4.9 | 27 | 89.8 | 23.3 | 8.6 | 25.9 | 13.9 |
| Inverse Probability weighting | | | | | | | | | | | |
| Treatment | 0.066*** | 0.068*** | 0.065*** | 0.068*** | 0.085*** | 0.315*** | 0.239*** | 0.125*** | 0.670*** | 0.125*** | 0.636*** |
| | (0.015) | (0.015) | (0.015) | (0.016) | (0.018) | (0.042) | (0.030) | (0.017) | (0.080) | (0.016) | (0.083) |
| With pre-analysis plan covariates | | | | | | | | | | | |
| Treatment | 0.083* | 0.085** | 0.084** | 0.091** | 0.108** | 0.299* | 0.231** | 0.113*** | 0.749** | 0.109*** | 0.628** |
| | (0.042) | (0.042) | (0.042) | (0.043) | (0.053) | (0.154) | (0.111) | (0.040) | (0.318) | (0.039) | (0.287) |
| Observations | 2,570 | 2,571 | 2,571 | 2,571 | 2,571 | 2,571 | 2,571 | 2,571 | 2,571 | 2,571 | 2,571 |
| Excluding wards affected by CAA-NRC protest | | | | | | | | | | | |
| Treatment | 0.061** | 0.063** | 0.059* | 0.059** | 0.083* | 0.341*** | 0.259*** | 0.136*** | 0.645*** | 0.118*** | 0.586** |
| | (0.030) | (0.030) | (0.030) | (0.028) | (0.042) | (0.104) | (0.086) | (0.035) | (0.197) | (0.035) | (0.231) |
| Observations | 2,461 | 2,461 | 2,461 | 2,461 | 2,461 | 2,461 | 2,461 | 2,461 | 2,461 | 2,461 | 2,461 |

Note: The marginal effects from the LPM/ OLS models are presented along with robust standard errors clustered at the ward level are presented in the parenthesis. *psacalc* command in STATA has been used to generate the bias adjusted treatment effect and *teffects* command to generate the estimates after inverse probability weighting. For IPW, robust standard errors are presented. * denotes significance at the 10%, ** at the 5%, and *** at the 1% levels.

**Appendix M: Mediation**

Table A9: Effects after controlling for EE and NE

| | No EE or NE controls | Controlling for EE | | | | | Controlling for NE | Controlling for EE and NE |
|---|---|---|---|---|---|---|---|---|
| *Last usage: toilet* | | | | | | | | |
| Treatment | 0.066** | 0.065** | 0.040 | 0.046* | 0.039 | 0.025 | 0.037 | 0.020 |
| | (0.029) | (0.028) | (0.028) | (0.027) | (0.025) | (0.025) | (0.027) | (0.025) |
| ***Controls additionally added in the regression model*** | | | | | | | | |
| How many out of 10 use a toilet? | No | Yes | No | No | No | Yes | No | Yes |
| How many out of 10 own a toilet? | No | No | Yes | No | No | Yes | No | Yes |
| Many people are constructing a toilet (agree) | No | No | No | Yes | No | Yes | No | Yes |
| Many people are using a toilet (agree) | No | No | No | No | Yes | Yes | No | Yes |
| How many out of 10 think one should use a toilet to defecate? | No | No | No | No | No | No | Yes | Yes |
| *Last two days primary defecation place: toilet* | | | | | | | | |
| Treatment | 0.067** | 0.066** | 0.041 | 0.048* | 0.039 | 0.025 | 0.037 | 0.020 |
| | (0.028) | (0.028) | (0.028) | (0.027) | (0.025) | (0.025) | (0.026) | (0.025) |
| ***Controls additionally added in the regression model*** | | | | | | | | |
| How many out of 10 use a toilet? | No | Yes | No | No | No | Yes | No | Yes |
| How many out of 10 own a toilet? | No | No | Yes | No | No | Yes | No | Yes |
| Many people are constructing a toilet (agree) | No | No | No | Yes | No | Yes | No | Yes |
| Many people are using a toilet (agree) | No | No | No | No | Yes | Yes | No | Yes |
| How many out of 10 think one should use a toilet to defecate? | No | No | No | No | No | No | Yes | Yes |
| *Exclusive toilets usage in last two days* | | | | | | | | |
| Treatment | 0.066** | 0.065** | 0.039 | 0.046* | 0.038 | 0.025 | 0.036 | 0.020 |
| | (0.029) | (0.028) | (0.028) | (0.027) | (0.025) | (0.025) | (0.027) | (0.025) |
| ***Controls additionally added in the regression model*** | | | | | | | | |
| How many out of 10 use a toilet? | No | Yes | No | No | No | Yes | No | Yes |

| How many out of 10 own a toilet? | No | No | Yes | No | No | Yes | No | Yes |
|---|---|---|---|---|---|---|---|---|
| Many people are constructing a toilet (agree) | No | No | No | Yes | No | Yes | No | Yes |
| Many people are using a toilet (agree) | No | No | No | No | Yes | Yes | No | Yes |
| How many out of 10 think one should use a toilet to defecate? | No | No | No | No | No | No | Yes | Yes |
| All other controls | Yes | Yes | Yes | Yes | Yes | Yes | Yes | Yes |
| Baseline mean at the ward-level | Yes | Yes | Yes | Yes | Yes | Yes | Yes | Yes |
| Observations | 2,562 | 2,562 | 2,562 | 2,562 | 2,562 | 2,562 | 2,562 | 2,562 |